\documentclass[traditabstract]{aa}

\usepackage{graphicx}
\usepackage{txfonts}
\usepackage{natbib}
\usepackage{longtable,lscape}

\bibpunct{(}{)}{;}{a}{}{,}

\newcommand{\feh}{\mathrm{[Fe/H]}}
\newcommand{\aFe}{\mathrm{[\alpha/Fe]}}
\newcommand{\teff}{T_\mathrm{eff}}
\newcommand{\logg}{\log g}

\begin{document}

\title{An absolutely calibrated $\teff$ scale from the Infrared Flux Method}
\subtitle{Dwarfs and subgiants}

\titlerunning{The effective temperature scale: resolving different versions}
\authorrunning{Casagrande et al.}

\author{L. Casagrande   \inst{1} \and
        I. Ram\'{\i}rez \inst{1} \and
	J. Mel\'endez   \inst{2} \and
        M. Bessell      \inst{3} \and
	M. Asplund      \inst{1}
       }

\institute{Max Planck Institute for Astrophysics,
           Postfach 1317, 85741 Garching, Germany \and
	   Centro de Astrof\'{\i}sica da Universidade do Porto, 
	   Rua das Estrelas 4150-762 Porto, Portugal \and
	   Research School of Astronomy and Astrophysics, 
	   Mount Stromlo Observatory, Cotter Rd, ACT 2611, Australia
          }

\date{Received; accepted}

\abstract{Various effective temperature scales have been proposed over the 
years. Despite much work and the high internal precision usually achieved, 
systematic differences of order $100$\,K (or more) among various scales are 
still present. We present an investigation based on the Infrared Flux Method 
aimed at assessing the source of such discrepancies and pin down their origin. 
We break the impasse among different scales by using a large set of solar 
twins, stars which are spectroscopically and photometrically identical to the 
Sun, to set the absolute zero point of the effective temperature scale to 
within few degrees. Our newly calibrated, accurate and precise temperature 
scale applies to dwarfs and subgiants, from super-solar metallicities to the 
most metal-poor 
stars currently known. At solar metallicities our results validate 
spectroscopic effective temperature scales, whereas
for $\feh \lesssim -2.5$ our 
temperatures are roughly $100$\,K hotter than those determined from model fits 
to the Balmer lines and $200$\,K hotter than those obtained from the excitation 
equilibrium of Fe lines. 
Empirical bolometric corrections and useful relations linking 
photometric indices to effective temperatures and angular diameters have been derived. 
Our results take full advantage of the high accuracy reached in absolute 
calibration in recent years and are further
validated by interferometric angular diameters and space based 
spectrophotometry over a wide range of effective temperatures and 
metallicities.}


\maketitle

\section{Introduction}

The determination of effective temperatures ($\teff$) in F, G and K
type stars has a long and notable history. Because of their long
lifetimes these stars retain in their atmospheres a fossil record of
the chemical elements in the interstellar medium at the time of their
formation. The stellar effective temperature is of
paramount importance for reliable abundance analyses and thus for
improving our understanding of Galactic chemical evolution.  

Stellar abundances are now routinely derived from high resolution
spectra, model atmospheres, and spectrum synthesis. While each of
these ingredients have their own issues regarding systematic
uncertainties, the dominant source of error is in many cases the
adopted $\teff$ of the star. Several indirect methods of $\teff$
determination have been devised to avoid the complications introduced
by the measurement of stellar angular diameters, which are necessary
to derive $\teff$ from basic principles
\citep[e.g.][]{hanbury74,vanBelle09}. Thus, most published values of
$\teff$ are model-dependent or based on empirical calibrations that
are not free from systematics themselves. 

It is therefore not surprising to find discrepancies among published
$\teff$ values. The ionization and excitation balance of iron lines in
a 1D LTE analysis is routinely used to derive effective temperatures
as well as $\logg$ and $\feh$. While for a sample of stars with
similar properties this method can yield highly precise relative
physical parameters \citep[][see Section \ref{cure} for its use on solar 
twins]{melendez09:twins,ramirez09}, non-LTE effects and departures 
from homogeneity can seriously
undermine effective temperature determinations, especially in
metal-poor stars \citep[e.g.][]{asplund05:review}. Similarly, the line-depth 
ratio technique has high internal precision, claiming to resolve temperature 
differences of order $10$\,K \citep[e.g.][]{gray91,gray94,kovtyukh03}
but it is not entirely model independent \citep[e.g.][]{caccin02,biazzo07} and
the uncertainty  on its zero point can be considerably large. Another
popular method for deriving $\teff$ in late-type stars is provided by
the study of the hydrogen Balmer lines, in particular H$\alpha$ and
H$\beta$ \citep[e.g.][]{nissen07,fuhrmann08}. For H lines
uncertainties related to observations and line broadening
\citep{barklem02}, non-LTE \citep{barklem07} and granulation effects
\citep[][]{asplund05:review,ludwig09,pereira09} all
influence the estimation of effective temperatures.  

In such a scenario, an almost model independent and elegant technique
for determining effective temperatures was introduced in the late 70's
by D.\ E.\ Blackwell and collaborators
\citep[][]{blackwell77,blackwell79,blackwell80} under the name of
InfraRed Flux Method (hereafter IRFM). Since then, a number of authors
have applied the IRFM to determine effective temperatures in stars with
different spectral types and metallicities
\citep[e.g.][]{bell89,alonso96:irfm,ramirez05a,casagrande06,gonzalez09}. The
main ingredient of the IRFM is infrared photometry, with the
homogeneous and all--sky coverage provided by 2MASS being the \emph{de
  facto} choice nowadays. As such, the IRFM can now be readily applied
to many stars, making it ideal to determine
colour--temperature--metallicity relations spanning a wide range of
parameters.  
The effective temperatures determined via IRFM are often regarded as a
standard benchmark for other techniques. Whilst they have high
internal accuracy and are essentially free from non-LTE and
granulation effects \citep[][Ramirez et al.~in prep.]{asplund01,casagrande09},
the reddening and absolute flux calibration adopted in such a
technique can easily introduce a systematic error as large as 100\,K
\citep{casagrande06}.  

The effective temperatures of dwarfs and subgiants are still heavily
debated with various $\teff$ scales behaving very differently
depending on colours and metallicities. One of the most critical
discrepancies occur at the metal-poor end,
for $\feh\lesssim-2.5$. In their work on the determination of
effective temperatures via IRFM, \citet{ramirez05a} found
temperatures significantly hotter than those previously published, in
particular those determined using the excitation equilibrium
method. Differences up to 500~K for the hottest ($\teff\simeq6500$\,K)
most metal-poor ($\feh\lesssim-3.0$) stars were reported
\citep[e.g.,][]{melendez04,melendez06a}. In this regime, the recent
IRFM investigation by \citet{gonzalez09} still supports a temperature
scale significantly hotter than excitation equilibrium and Balmer
lines, but $\sim 90$~K cooler than \citet{ramirez05a}. 

The abundance pattern measured in metal-poor stars is important for
our quest to understand Galactic chemical evolution and Big Bang
nucleosynthesis: two notable examples are the oxygen abundance and the
lithium trend with metallicity, both of which crucially depend on
the adopted $\teff$ scale. For example, a change of $+100$\,K in 
$\teff$ would decrease the [O/Fe] ratio in turn-off metal-poor stars by 
$\sim$0.08 dex when using the OI triplet and FeII lines \citep{melendez06b}, 
while the same change in $\teff$ would increase the Li abundance by $\sim$ 0.07 
dex \citep[e.g.][]{melendez04,melendez09:review,melendez09:lithium}.

At higher metallicities, which encompass most of the stars in the
solar neighbourhood, the situation is also uncertain, with
spectroscopic  effective temperatures in rough agreement with the IRFM
scale of \citet{casagrande06}. The latter is then about 100~K hotter
than the IRFM temperatures of \citet{ramirez05b} whilst the recent
implementation of  \citet{gonzalez09} falls in between these two
extremes. These differences are somewhat puzzling considering that all
recent works on the IRFM have used 2MASS photometry. Effective temperature
calibrations are also crucial in the context of deriving reliable
colours for theoretical stellar models, which apart from few notable
exceptions \citep[e.g.][]{vandenberg03} have to resort entirely to theoretical
flux libraries. 
  
The aim of this work is to uncover the reason(s) behind such a
confusing scenario and provide a solution to different IRFM effective
temperature scales currently available in literature. As we discuss
throughout the paper, this ambitious task is accomplished by using
solar twins which allow us to set the absolute zero point of the
$\teff$ scale. This result is further validated using interferometric
angular diameters and space-based spectrophotometry. 

The paper is organized as follows. In Section \ref{compare} we compare
the results obtained from different authors, focusing in particular on
two independent implementations of the IRFM \citep{ramirez05a,
  casagrande06} when the same input data are used. This approach
allows us to precisely identify where different $\teff$ scales
originate from. A cure to such an impasse is then provided in Section
\ref{cure}. The validation of our results, together with the new both
precise and accurate effective temperature scale are presented in
Section \ref{valid} to \ref{colte}. We finally conclude in Section
\ref{conclusions}.  

\section{Comparing different versions}\label{compare}

\begin{figure*}
\centering
\includegraphics[scale=0.4]{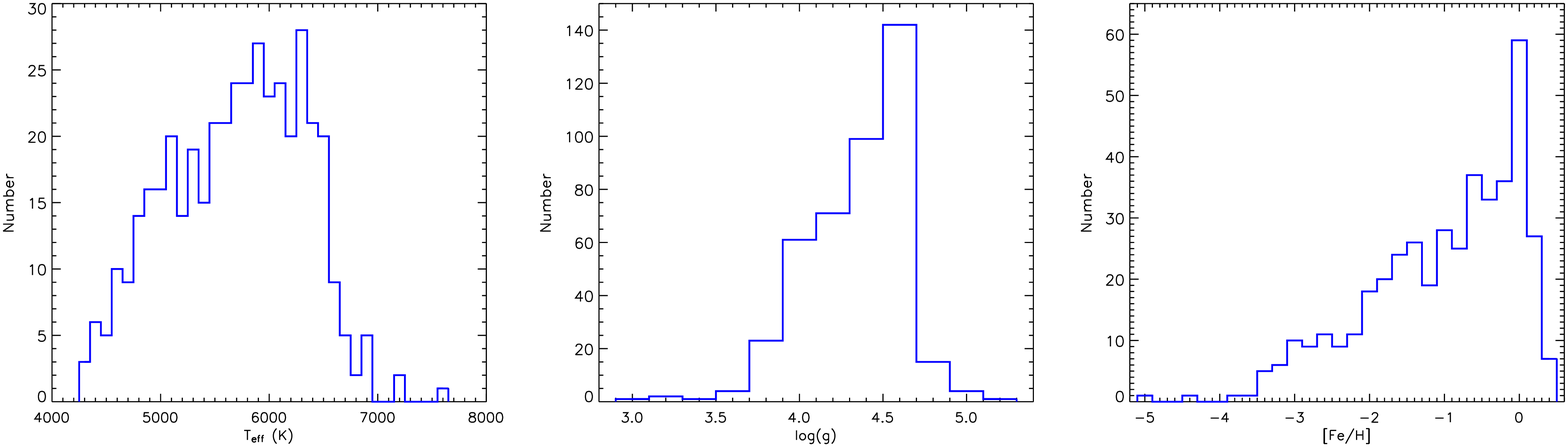}
\caption{Distribution of $\teff$, $\logg$ and $\feh$ for the 423 stars in 
our sample.}
\label{f:sample}
\end{figure*}

In this paper we use an updated version the IRFM implementation
described in \citet{casagrande06} to nail down the reasons behind
different $\teff$ scales. Our implementation works in the 2MASS system
and fully exploits its high internal consistency thus making it well
suited to the purpose of the present investigation. The core of the
present study is to carry out a detailed comparison with the
\citet{ramirez05a} implementation when the same input data are
used. For the sake of precision, notice that hereafter, when we refer 
to a $\teff$ determined by \citet{ramirez05a} we are referring to the 
effective temperatures
determined using that implementation and not the original values given
in that paper. This is because of the updated (and more consistent)
input data used here and also because some of the stars presented in
this work do not have IRFM $\teff$ values published yet. In fact, in
order to reveal trends with metallicity and/or effective temperature,
our sample is specifically built to cover as wide a range as possible
in those parameters (Figure \ref{f:sample}).

\subsection{Input sample}\label{sampleJC}

The main ingredient of the IRFM is optical and infrared
photometry. The technique depends very mildly on other stellar
parameters, such as metallicity and surface gravity, which are needed
to interpolate on a grid of model atmospheres (see Section
\ref{proscons}). Below we present the papers from which we gathered
$\feh$ and $\logg$ for all our stars and we also give references to
the photometric sources.  

The metal-rich dwarfs come from \citet{casagrande06} who also provide
homogeneous and accurate $BV(RI)_C$ photometry while additional
metal-rich dwarfs and subgiants are from \citet{ramirez05a}. We
complement the sample with a number of moderately metal-poor stars
from the study of \citet{fabbian09} and metal-poor turn off stars from
\citet{hosford09}. To investigate the metal-poor end of the $\teff$
scale in more detail, stars with reliable input data from
\citet{ramirez05a}, \citet{bonifacio07} and \citet{aoki09} were
added. Finally, to explore for the first time the hyper-metal-poor
regime via IRFM the subgiants HE0233-0343 \citep[$\feh \lesssim
  -4$][]{ana08} and HE1327-2326 \citep[$\feh \le
  -5$][]{frebel05:nature,aoki06,frebel08,korn09} were included. 

New $UBV(RI)_C$ photometric observations for some of the metal-poor
stars in the aforementioned papers were conducted by Shobbrook \& Bessell 
(1999; private communication) and are given in Table \ref{newUBVRI}. For the
remaining stars, optical Johnson-Cousins photometry was taken either
from \citet{beers07} or the General Catalogue of Photometric Data
\citep{mermilliod97}.  

\begin{table}
\centering
\caption{New Johnson-Cousins photoelectric photometry obtained for some of the metal-poor stars in the sample. Each measurement comprises an average of 4 observations per star. The rms of individual observations are $0.02$ for the V magnitude, $0.015$ for the U-B colour and $0.008$ mags for B-V, V-R, R-I, V-I colours.}
\label{newUBVRI}
\begin{tabular}{lccccc}
\hline\hline
Name	     &        $U$    &    $B$  &  $V$   & $R_C$  &   $I_C$ \\     
\hline
HD3567       &        9.556  &   9.695 &  9.240 &  8.941 &  8.631  \\ 
HD16031      &       10.004  &  10.197 &  9.770 &  9.484 &  9.184  \\ 
HD19445	     &        8.207  &   8.503 &  8.026 &  7.737 &  7.394  \\ 
HD34328	     &        9.683  &   9.903 &  9.416 &  9.106 &  8.773  \\ 
HD45282      &        8.659  &   8.672 &  8.010 &  7.610 &  7.196  \\
HD59392	     &       10.048  &  10.217 &  9.761 &  9.457 &  9.142  \\ 
HD64090	     &        8.762  &   8.951 &  8.295 &  7.935 &  7.536  \\ 
HD64606      &        8.277  &   8.140 &  7.412 &  6.994 &  6.561  \\
HD74000      &        9.880  &  10.071 &  9.656 &  9.381 &  9.080  \\ 
HD84937      &        8.485  &   8.702 &  8.306 &  8.047 &  7.759  \\ 
HD94028	     &        8.421  &   8.640 &  8.202 &  7.917 &  7.585  \\ 
HD102200     &        9.009  &   9.189 &  8.739 &  8.449 &  8.141  \\ 
HD106038     &       10.431  &  10.627 & 10.153 &  9.857 &  9.529  \\ 
HD108177     &        9.874  &  10.082 &  9.647 &  9.362 &  9.052  \\ 
HD110621     &       10.230  &  10.385 &  9.932 &  9.628 &  9.313  \\ 
HD114762     &        7.738  &   7.833 &  7.283 &  6.967 &  6.629  \\
HD116064     &        9.099  &   9.282 &  8.833 &  8.520 &  8.189  \\ 
HD122196     &        9.055  &   9.212 &  8.753 &  8.444 &  8.112  \\
HD132475     &        8.983  &   9.100 &  8.563 &  8.216 &  7.855  \\
HD134169     &        8.115  &   8.193 &  7.663 &  7.342 &  7.011  \\
HD134439     &       10.033  &   9.881 &  9.118 &  8.661 &  8.220  \\ 
HD140283     &        7.502  &   7.692 &  7.205 &  6.876 &  6.522  \\ 
HD160617     &        9.014  &   9.188 &  8.740 &  8.431 &  8.108  \\ 
HD163810     &       10.185  &  10.272 &  9.660 &  9.280 &  8.897  \\
HD179626     &        9.601  &   9.710 &  9.188 &  8.849 &  8.502  \\ 
HD181743     &        9.911  &  10.140 &  9.683 &  9.375 &  9.062  \\ 
HD188510     &        9.303  &   9.452 &  8.851 &  8.486 &  8.100  \\ 
HD189558     &        8.214  &   8.299 &  7.740 &  7.392 &  7.034  \\ 
HD193901     &        9.049  &   9.183 &  8.644 &  8.307 &  7.964  \\ 
HD194598     &        8.666  &   8.844 &  8.356 &  8.055 &  7.739  \\ 
HD199289     &        8.660  &   8.803 &  8.287 &  7.972 &  7.643  \\
HD201891     &        7.740  &   7.908 &  7.390 &  7.081 &  6.737  \\ 
HD213657     &        9.869  &  10.063 &  9.646 &  9.368 &  9.068  \\ 
HD215801     &       10.272  &  10.471 & 10.038 &  9.732 &  9.418  \\ 
HD219617     &        8.425  &   8.621 &  8.153 &  7.845 &  7.525  \\ 
HD284248     &        9.407  &   9.650 &  9.208 &  8.927 &  8.608  \\ 
HD298986     &       10.316  &  10.506 & 10.062 &  9.774 &  9.470  \\ 
BD+17 4708   &        9.718  &   9.922 &  9.476 &  9.183 &  8.854  \\ 
BD+02 3375   &       10.174  &  10.414 &  9.944 &  9.635 &  9.297  \\ 
BD-04 3208   &       10.203  &  10.375 &  9.977 &  9.709 &  9.417  \\ 
BD-13 3442   &       10.529  &  10.655 & 10.266 &  9.994 &  9.704  \\ 
CD-30 18140  &       10.155  &  10.365 &  9.946 &  9.663 &  9.353  \\ 
CD-33 3337   &        9.436  &   9.581 &  9.109 &  8.814 &  8.490  \\ 
\hline
\end{tabular}
\end{table}

Infrared $JHK_S$ photometry for the entire sample is available from
the 2MASS catalogue \citep{skr06} which also includes the uncertainty
for each observed magnitude (``j\_'', ``h\_'' and ``k\_msigcom''). 
The infrared median total photometric error of our sample is 
0.07 mag.~(i.e.\ ``j\_''$+$``h\_''$+$``k\_msigcom''$=0.07$) and 
never exceeds $0.14$ mag. Such an
accuracy in the infrared photometry implies a mean (maximum) internal
error in $\teff$ of 25~K (50~K). Notice that the effective internal
accuracy is slightly worse because of additional uncertainties stemming from
the optical photometry, $\feh$ and $\logg$. Altogether our final
sample consists of 423 stars: all have $BVJHK_S$ photometry while more than 
half have also $(RI)_C$ magnitudes available\footnote{Other than 
being available only for a limited number of stars, we did not 
use $U$ magnitudes because of the little flux emitted in this 
region and the high uncertainties related to the absolute 
calibration and standardization of this passband in both 
observed and synthetic photometry \citep[e.g.][and references 
therein]{bessell05}.}. 

Proper reddening corrections are crucial to determine $\teff$ via
IRFM. We have tested that 0.01 mag.~in $E(B-V)$ translates into an IRFM
effective temperature roughly 50~K hotter. 
Reddening is usually zero for
stars lying within the local bubble $\lesssim 70$~pc from the Sun 
\citep[e.g.][]{leroy93,lallement03} and so we have adopted 
$E(B-V)=0$ for all stars having {\it Hipparcos} parallaxes 
\citep{vanLeeuwen07} and satisfying this 
requirement on the distance. For the remaining stars we updated
the reddening corrections in \citet{ramirez05a} based on various
extinction maps and, in particular for metal-poor stars when
archive high resolution spectra were available, using interstellar
NaD absorption lines \citet{melendez09:lithium}.
In broad-band photometry the definition of the effective wavelength of
a filter ($\lambda_{\rm{eff}}$) shifts with the colour of the star
\citep[e.g.][]{bessell98,casagrande06}. Therefore a given $E(B-V)$
colour excess must be scaled according to the intrinsic colour of the
source under investigation. From the reddening $E(B-V)$, we computed
the extinction in each band adopting the reddening law of
\citet{donnell94} for the optical and \citet{cardelli89} for the
infrared, using the improved estimation of the stellar intrinsic flux
obtained at each iteration to bootstrap the computation of the correct
$\lambda_{\rm{eff}}$ in our IRFM code.  
  
\subsection{The IRFM: pros and cons}\label{proscons}

\begin{figure}
\centering
\includegraphics[scale=0.44]{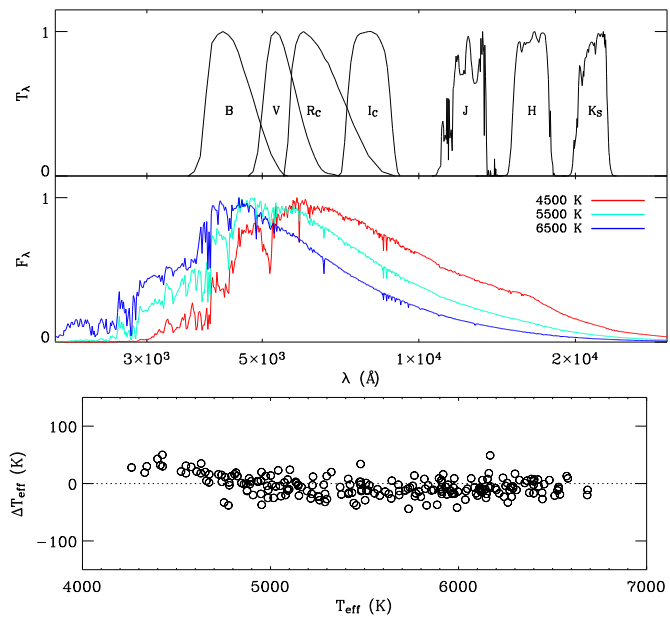}
\caption{Top panel: Johnson-Cousins-2MASS filter sets used in this
  work. Middle panel: synthetic solar metallicity spectra at different
  $\teff$. For the sake of comparison all curves have been normalized
  to unit. Bottom panel: difference in effective temperatures
  $\rm{with} - \rm{without}$ using $(RI)_C$ magnitudes to recover the
  bolometric flux.}
\label{f:sed}
\end{figure}

The basic idea of the IRFM is to compare the ratio between the
bolometric flux $\mathcal{F}_{Bol}\rm{(Earth)}$ and the infrared
monochromatic flux $\mathcal{F}_{\lambda_{IR}}\rm{(Earth)}$, both
measured at the top of Earth's atmosphere (the so-called observational
$R_{obs}$ factor) to the ratio between the surface bolometric flux
($\sigma \teff^4$) and the surface infrared monochromatic flux
$\mathcal{F}_{\lambda_{IR}}(\teff,\feh,\logg)$ determined
theoretically for any given set of stellar parameters. The latter is
called the theoretical $R_{theo}$ factor. For stars hotter than about
$4200$~K, infrared photometry longward of $\sim 1.2\,\mu$m ensures we
are working in the Rayleigh-Jeans part of a stellar spectral energy
distribution, a region largely dominated by the continuum which
linearly depends on $\teff$ and thus only mildly on model atmospheres
(Figure \ref{f:sed}). An extension of the technique to cooler
effective temperatures using near-infrared photometry is possible, as
shown by \citet{casagrande08}, but this is outside the purpose of the
present paper. 

$R_{obs}$ and $R_{theo}$ can be immediately rearranged to determine
$\teff$, effectively reducing the entire problem to properly recover
$\mathcal{F}_{Bol}\rm{(Earth)}$ and
$\mathcal{F}_{\lambda_{IR}}\rm{(Earth)}$. Both quantities are
determined from photometric observations, but an iterative procedure
is adopted to cope with the mildly model dependent nature of the
bolometric correction. In our case we use the fluxes predicted by the 
\citet{castelli04} grid
of model atmospheres starting with an initial estimate of the
effective temperature and interpolating at the appropriate $\feh$ and
$\logg$ until convergence in $\teff$ is reached within $1$\,K. By
doing so, we also obtain a synthetic spectrum tailored to the
effective temperature empirically determined via IRFM. 

Though we interpolate at the proper $\feh$ and $\logg$ of each star,
the dependence of the technique on such parameters is minor
\citep[e.g.][]{ramirez05a,casagrande06}. This feature makes the IRFM
superior to any spectroscopic methods to determine $\teff$ --provided
the reddening is accurately known-- since in the latter the effects of
$\teff$, $\logg$ and $\feh$ are usually strongly coupled and the model
dependence is much more important.  

The errors are estimated using realistic observational uncertainties
in a Monte Carlo simulation plus the systematics arising from the
adopted absolute calibration, as described in
\citet{casagrande06}. With the improved absolute calibration used in
this paper, systematics amount to $15$\,K in $\teff$ and $0.3$\% in
bolometric flux (Section \ref{tuning}). For stars approximately cooler
than $5000$\,K, $(RI)_C$ photometry is crucial to properly compute the
bolometric flux. This can be appreciated in the lower panel of Figure
\ref{f:sed}: below this temperature a trend appears using $BVJHK_S$
magnitudes only. Missing the peak of the energy distribution clearly
leads one to underestimate the bolometric flux thus returning cooler
effective temperatures. We have linearly fitted the trend below
$5000$\,K to remove such differences in both $\teff$ and
$\mathcal{F}_{Bol}$ when $(RI)_C$ photometry was not available. For
$\teff > 5000$\,K no obvious trend appears: constant offsets of merely
$7$\,K in $\teff$ and $0.15$\% in bolometric flux have been found,
consistent with the effect that the absolute calibration in $(RI)_C$
can introduce. For the sake of homogenizing the stellar parameters
derived in this work, also these small offsets have been corrected for
stars with no $(RI)_C$ photometry. 

The effective temperature can be determined from any infrared
photometric band, in our case $JHK_S$ from 2MASS. Ideally all bands
should return the same $\teff$, but photometric errors and zero point
uncertainties in the absolute calibration of each band introduce
random plus systematic differences. In the case of 2MASS, those amount
to few tens of K as we show later. 

The magnitude in a given band $\zeta$ is converted into a physical
flux (i.e.\ $\rm{erg\,cm^{-2}\,s^{-1}}\,\AA^{-1}$) via 
\begin{equation}\label{fluxes} 
\mathcal{F}_{\zeta}\rm{(Earth)}=\mathcal{F}_{\zeta}^{std}\rm{(Earth)} 10^{-0.4(m_\zeta-m_\zeta^{std})}
\end{equation}
which depends on the zero point ($m_\zeta^{\rm std}$) and the absolute
flux calibration ($\mathcal{F}_{\zeta}^{\rm std}$) of the standard
star defining the photometric system under use\footnote{We point out
  that Eq.~(\ref{fluxes}) holds exactly for a heterochromatic
  measurement, while for computing a monochromatic flux from the
  observed photometry, an additional correction (the so called
  $q$-factor) must be introduced to account for the fact that the zero
  point of the photometric system is defined by a standard star, which
  usually has a different spectral energy distribution across the
  filter window with respect to the problem star
  \citep[e.g.][]{alonso96:irfm,casagrande06}.}.  

Most of the photometric systems, including Johnson-Cousins and 2MASS,
use Vega as the zero point standard. Vega's flux and magnitudes in
different bands have been notoriously difficult to measure with
sufficient accuracy \citep[e.g.][and references therein]{gray07}. The
problem is only apparently resolved when resorting to $R_{obs}$: in
the ideal case of a unique template spectrum for Vega the choice of
its absolute calibration would cancel out in the ratio. In practice,
the situation is far from this since the pole-on and rapidly rotating
nature of this star imposes the use of a composite absolute calibrated
spectrum for different wavelength regions \citep[e.g.][and references
  therein]{casagrande06}. Such complication does not disqualify Vega
as a spectrophotometric standard, but it makes its use more problematic. 
From Eq.~(\ref{fluxes}) it can be immediately noticed that a change of
0.01 mag.~corresponds to a change of about 1\% in flux. Since it is
possible to interchangeably operate on both zero points and fluxes,
for the sake of our discussion it is their composite effect that must
be considered, though in the following we shall usually refer to
fluxes.  

Recently, HST spectrophotometry for Vega has provided a unique
calibrated spectrum extending from 3200 to 10000 \AA\, with $1-2$\%
accuracy \citep{bohlin07}. In the infrared, once the zero points newly
determined from \citet{apellaniz07} are used, this result is also in
broad agreement with the 2MASS absolute calibration provided by
\citet{cohen03}. \citet{rieke08} have also recently reviewed the
absolute physical calibration in the infrared, substantially
validating the accuracy of 2MASS: their recommended 2\% increase of
flux in $K_S$ band is in fact compensated by their newly determined
zero point for Vega, thus implying an effective change in the overall
$K_S$ calibration of only $0.2$\%. We have tested all these different
possibilities; with respect to the HST and 2MASS calibration adopted
in \citet{casagrande06} the derived $\teff$ are affected at most by
$20$\,K. Such difference is thus within the aforementioned global
$2$\% uncertainty which allows for systematics in $\teff$ of order
$40$\,K. Our zero points and absolute fluxes are essentially identical
to those adopted in \citet{casagrande06} except for a small
fine-tuning which will be further discussed in Section \ref{cure}. 

Despite the recent increasing concordance in establishing absolute
fluxes, the uncertainties which have historically plagued Vega are
crucial in the context of understanding the effective temperatures
determined via IRFM by various authors. We have tested that uncorrelated 
changes of a few percent in the absolute calibration of optical bands (needed to
recover the bolometric flux) can introduce spurious trends with
$\teff$ and $\feh$ up to few tens of K. Similar changes in the 
absolute calibration of infrared bands have only minor impact 
on the bolometric flux, but as already mentioned, $\teff$ is very 
sensitive to them since they enter explicitly in the definition 
of $R_{obs}$: increasing all of them by
2\% translates into a decrease of approximately 40~K in
$\teff$. Considering that differences of few percent in the adopted
zero points and fluxes are commonly present among various IRFM
implementations, it can be immediately realized that they are responsible
for systematic differences among various authors.  

\subsection{Alonso et al. (1996) scale}\label{comparisonA}

\begin{figure}
\centering
\includegraphics[scale=0.44]{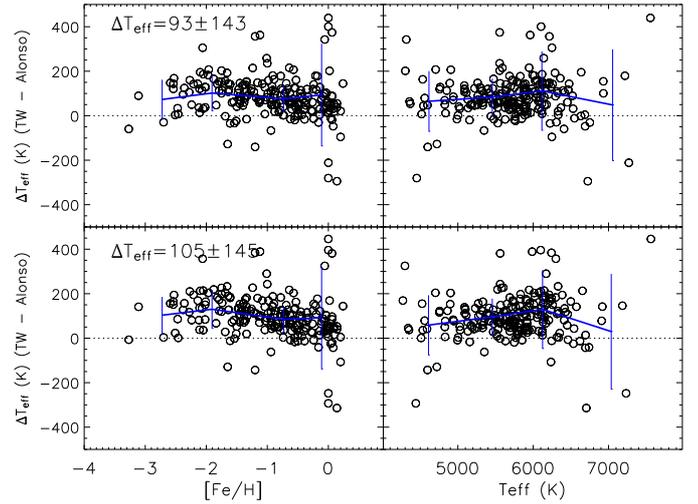}
\caption{Difference between the effective temperatures obtained in this work (TW) and those reported in \citet{alonso96:irfm} for 220 stars in common. In case of reddening, only stars with values of $E(B-V)$ equal to within 0.02 mag.~have been plotted. Thick continuous lines connect the means computed in equally spaced bins of $\feh$ and $\teff$. Error bars are the standard deviation in each bin. Top panels: when \citet{kurucz93:cd13} models are used in our version of the IRFM. Bottom panels: when the new \citet{castelli04} models are used instead. Below $\feh=-1.5$ the new models support $\teff$ hotter by 20 to 40~K.}
\label{f:alonso}
\end{figure}

One of the most extensive applications of the IRFM to Pop I and II
dwarfs is that of \citet{alonso96:irfm}, which was based on the
infrared photometry collected at the TCS 
\citep[Telescopio Carlos Sanchez,][]{alonso94:photometry}
and absolutely calibrated using a semi-empirical approach 
relying on (mostly) giant stars with measured angular diameters to 
determine the reference absolute fluxes \citep{alonso94:absolute_calibration}. 
The comparison between our
$\teff$ and those by \citet{alonso96:irfm} is shown in Figure
\ref{f:alonso}. Despite the scatter arising from the different
input data we used, there is a clear offset with our scale being
systematically hotter. No obvious trends in $\teff$ and $\feh$
appear. This offset is easily explained in terms of the absolute
calibration underlying the two different photometric systems
adopted. This involves the transformation from TCS to 2MASS system
\citep[see also the discussion in][]{casagrande06}, which could in
principle introduce additional noise (see Section \ref{comparisonRM}). 
A more detailed description of the absolute calibration (and 
angular diameters) employed by Alonso and a comparison with our own 
is presented in Appendix A.

\begin{figure*}
\centering
\includegraphics[width=0.99\textwidth]{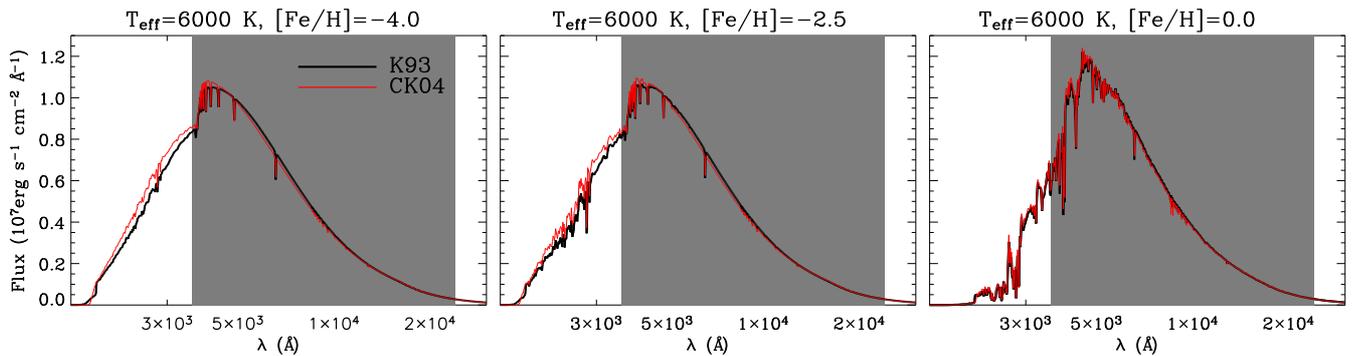}
\caption{Comparison between \citet{kurucz93:cd13} (thick line) and 
\citet{castelli04} (thin line) synthetic spectra at different metallicities 
for an assumed $\logg=4.0$.
Shaded area is the wavelength region covered by our multiband photometry. The 
difference in the UV flux gets more prominent when going to more metal-poor 
stars, but for the sake of the IRFM is entirely negligible at solar 
metallicity.}
\label{f:k93}
\end{figure*}

An area of particular interest is the determination of effective
temperatures in very metal-poor, turn-off stars. We have tested the
effect of using the new \citet{castelli04} model atmospheres in the
IRFM instead of the \citet{kurucz93:cd13} adopted by
\citet{alonso96:irfm}. The IRFM is known to be little model 
dependent \citep[e.g.][]{asplund01,casagrande09} and in fact there 
are no big differences except at the 
lowest metallicities, where \citet{castelli04} support effective 
temperatures hotter by $\sim40$~K. The reason for such a discrepancy 
stems from the new models returning higher flux below 
$\sim 4000$~\AA, a region where the most metal-poor, turn-off 
stars commence emitting non negligible amounts of energy. Since 
we do not have UV photometry (and its standardization would be uncertain), 
we must rely on model atmospheres to determine the flux over this region 
(Figure \ref{f:k93}). The latest model atmosphere calculations show 
excellent agreement as we checked that nearly identical $\teff$ 
are obtained when the new MARCS models \citep{gustafsson08} are 
used instead of those by \citet{castelli04} (also Section \ref{he1327}), but 
see \citet{edvardsson08} for a discussion of the performance of model 
atmospheres in the blue and ultraviolet.

\subsection{Ram\'irez \& Mel\'endez (2005) scale}\label{comparisonRM}

A revision of the \citet{alonso96:irfm} implementation of the IRFM was
carried out by \citet{ramirez05a} based on the TCS (for the
computation of $R_{theo}$) and Johnson's (for the computation of the
bolometric fluxes) \textit{JHK} photometric systems
\citep{alonso94:photometry,bessell88}. Here we replicate the $\teff$
determination by \citet{ramirez05a} for comparison purposes. When running their implementation, we transformed the 2MASS photometry into TCS using their equations. However, when comparing the transformed and original \textit{JHK} values for these stars we found zero point differences at the level of 0.01 magnitudes: these offsets are within the photometric uncertainties and smaller than the scatter in the fits leading to the transformation equations, but they introduce changes in the derived $\teff$ values up to few tens of K (see Section \ref{proscons}). Therefore we took those into account to precisely transform 2MASS data into the TCS system.  

The \citet{ramirez05a} bolometric fluxes were determined using the K-band bolometric correction calibration by \cite{alonso95}, which depends only on the Johnson $(V-K)$ colour index and the stellar metallicity.\footnote{We have also tested that in the context of computing bolometric fluxes for this work, the updated J.~Carpenter transformations from 2MASS to Johnson available online at: http://www.astro.caltech.edu/$\sim$jmc/2mass/v3/transformations are instead accurate enough and insensitive to small zero point changes.}
This calibration is internally accurate within its ranges of applicability and one would expect that extrapolations slightly outside these ranges would still provide reliable results at low metallicities. This approach was followed by \citet{ramirez05a}.
With regards to the absolute flux calibration in the infrared, \citet{ramirez05a} adopted that of \cite{alonso94:absolute_calibration}, which is valid for TCS $JHK$ photometry while we use an update of \cite{cohen03} for the $JHK_S$ 2MASS system (see also Section \ref{cure}). 

\begin{figure}
\includegraphics[scale=0.44]{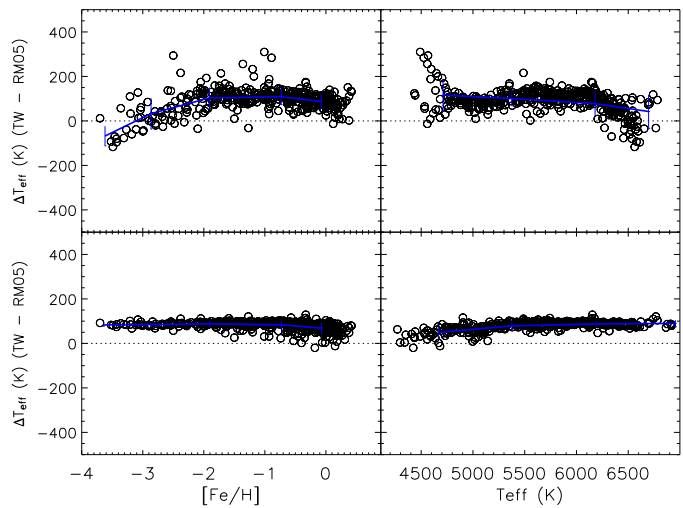}
\caption{Top panels: difference between the effective temperatures of this work (TW) and those obtained when the same input data are used in the \citet{ramirez05a} implementation (RM05). Bottom panels: as in the top panels but for the \citet[][]{ramirez05a} temperatures re-determined using the bolometric fluxes obtained in this work.}
\label{f:rm05}
\end{figure}

The difference between our results and \citet{ramirez05a} when the same input data and reddening values are adopted is illustrated in the top panels of Fig.~\ref{f:rm05}. Some of the scatter arise from transforming 2MASS magnitudes into TCS, but clear trends with both with $\teff$ and $\feh$ are present. For the bulk of the stars with $\feh > -2.0$ and $4800< \teff < 6200$\,K a roughly constant offset of about 100\,K is observed, our stars being hotter. In the metal-rich regime such an offset is present also for hotter stars ($\teff > 6200$\,K), but reduces somewhat for the coolest metal-rich dwarfs, reaching a minimum of about 50\,K at $\teff\simeq4500$\,K. A steep trend is seen for moderately metal-poor dwarfs ($-2.0<\feh<-1.0$) below $4800$~K, a region with few or no calibrating stars in \citet{alonso95}. 
For the warmer, most metal-poor stars in the sample, the differences decrease sharply with increasing $\teff$ and decreasing $\feh$, quickly becoming negative i.e., \citet{ramirez05a} temperatures become warmer, reaching a maximum value of about $-100$\,K at $\teff\simeq6500$\,K and $\feh\simeq-3.5$. 

To investigate the source of these differences, we re-calculated the IRFM temperatures of \citet{ramirez05a} using our bolometric fluxes instead of the calibration formulae adopted by \citet{ramirez05a}. This choice is perfectly legitimate, since what is crucial in the IRFM are the infrared fluxes which appear explicitly in the definition of $R_{obs}$, while $\teff$ depends only mildly on the bolometric flux (Section \ref{proscons}). Therefore, adopting our bolometric fluxes is substantially independent of the underlying temperature scale, i.e.\ the \citet{ramirez05a} scale is still recovered despite now using the new bolometric fluxes determined in the present work. The result of this exercise is shown in the bottom panels of Fig.~\ref{f:rm05}. The major trends caused from extrapolating the \citet{alonso95} bolometric formulae now disappear with a constant offset $\Delta \teff=85 \pm 13$\,K above $5000$\,K. The small trend that remains below this temperature corresponds to the threshold where \citet{ramirez05a} stop using the $J$ band to determine $\teff$, which in the TCS system usually returns slightly cooler $\teff$ than $H$ and $K$ bands. 

From this comparison it is clear that \citet{ramirez05a} temperatures for the metal-poor turn-off stars are warmer due to the use of a photometric calibration to derive the bolometric fluxes. In fact, we realize that the \cite{alonso95} formula is robust down to $\feh\simeq-2.5$ and up to $\teff\simeq6500$\,K but only a few calibrating stars more metal-poor or warmer exist in their sample. \citet{ramirez05a} use of this formula in regions where the calibration is uncertain (and in some cases outside of the ranges of applicability) has resulted in the very high temperatures of the more metal-poor turn-off stars. The extrapolation is, of course, not a valid procedure, even though one might expect the $\feh$ dependence of the calibration not to be so important at low metallicity. However, as can be seen from Fig.~4 in \cite{alonso95}, at these relatively high temperatures, the effect of $\feh$ is very important and such extrapolations should not be performed. 

The difference that remains after adopting consistent bolometric fluxes between this work and \citet{ramirez05a} (lower panels of Fig.~\ref{f:rm05}) is mostly due to the use of different infrared absolute flux calibrations. 
In fact, by lowering the absolute fluxes adopted by \citet{ramirez05a} by about 4\%, the mean difference reduces to almost zero. 
We thus conclude that our and \citet{ramirez05a} IRFM implementations can be made perfectly compatible if the same input parameters and flux calibration are used. 

\subsection{Gonz\'alez Hern\'andez \& Bonifacio (2009) scale}\label{comparisonGB}

\begin{figure}
\centering
\includegraphics[scale=0.44]{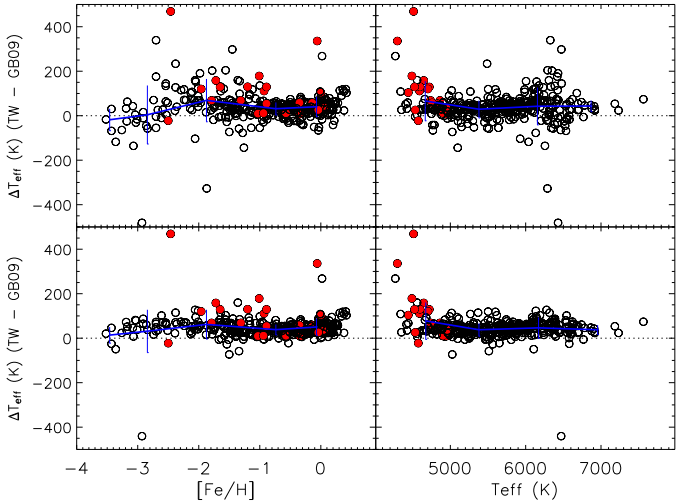}
\caption{Top panels: difference between the effective temperatures of this work (TW) and those in \citet{gonzalez09} (GB09) for 380 stars in common. Filled circles are stars with $\teff < 5000$\,K without $(RI)_C$ photometry in GB09. Bottom panels: as in the top panels, but when the same reddening corrections are used.}
\label{f:gb09}
\end{figure}

The most recent work on the IRFM is that by \citet{gonzalez09}, which
is also based on 2MASS photometry. The main difference between theirs
and our implementation is the different absolute calibration and zero
points adopted for Vega. They based their work on the \citet{castelli94} model and \citet{McCall04} magnitudes instead of the HST \citep{bohlin04a,bohlin07} and 2MASS \citep{cohen03} values that we use. Although such differences are within the current observational errors, in the infrared the combined effect of their fluxes and zero points is on average $1.5-2.0$\% higher than ours, implying effective temperatures cooler by $30-40$\,K (see Appendix A). This can be immediately appreciated in Figure \ref{f:gb09}, which indeed shows a constant offset of this magnitude for stars in common, thus confirming the offset noticed by \citet{gonzalez09} for stars in common with \citet{casagrande06}. 

The very steep trend at the lowest metallicities is due to the
different reddening corrections we adopt with respect to theirs. When
the same $E(B-V)$ values are adopted (bottom panels in Figure
\ref{f:gb09}), the offset remains constant throughout the entire
$\feh$ and $\teff$ range, except for few outliers due to the different
input data (mostly optical photometry) adopted. This clearly stresses
the importance of proper reddening correction for determining
effective temperatures via IRFM in stars outside of the local
bubble. For the most metal-poor stars in the sample,
we use interstellar NaD lines to achieve higher precision (Section
\ref{sampleJC}) while \citet{gonzalez09} resorted to reddening maps
scaled by the distance and the galactic latitude of the star and scale
height of the dust layer. The trend towards cooler effective
temperatures that we obtain in this regime thus stem entirely from better
reddening corrections. Finally, we suspect that the trend for $\teff < 5000$\,K is due to the absence of $(RI)_C$ colours in \citet{gonzalez09} 
(Section \ref{proscons}, bottom panel of Figure \ref{f:sed}).

\section{Resolving different versions}\label{cure}

It is clear from the discussion above that we now understand where different $\teff$ scales originate from and the crucial role played by the absolute calibration. Our approach has been to adopt the latest calibration available for each photometric system: currently those are accurate at the $2$\% level, implying possible systematic uncertainties of order $40$\,K. 
Here we want to improve upon this uncertainty using an independent verification of the absolute calibration adopted. 

\subsection{Solar twins}\label{twins}

The use of solar-type stars to calibrate photometric systems has a long and noble history, which relies on taking absolutely calibrated measurements of the Sun and computing synthetic colours to compare with other solar-type stars \citep[e.g.][]{johnson65,campins85,rieke08}. This rationale can be extended to other physical properties, namely using the solar effective temperature $T_{\rm{eff},\odot}=5777$\,K as the average value for solar-type stars \citep[e.g.][]{masana06}. This technique is well established and goes under the name of solar analogs method, but there is some sort of {\it petitio principii} in the underlying $\teff$ scale adopted and/or the solar colours assumed to select solar analogs in first instance.   

A way to break such a degeneracy is provided by solar twins, i.e.~stars with 
spectra indistinguishable from the Sun \citep[][]{cayrel89,demello97}. 
Our twins were drawn from an initial sample of about 100 stars broadly
selected to be solar like: the identification of the best ones was
based on a strictly differential analysis of high-resolution ($R \sim
60000$) and high signal-to-noise ($S/N \gtrsim 150$) spectra with
respect to the solar one reflected from an asteroid and observed with
the same instrument. Within this initial sample, the selection
criterion adopted to identify the best twins did not assume any {\it a priori} 
effective temperature or colour, but was based on the measured relative 
difference in equivalent widths and equivalent widths vs.\ excitation 
potential relations with respect to the observed solar reference 
spectrum and thus entirely model independent \citep[]{melendez06b,melendez07}. 
Since the spectra of the solar twins match so closely the solar one, 
exceedingly accurate differential spectroscopic analysis with respect to 
$T_{\rm{eff},\odot}, \rm{[Fe/H]}_{\odot}$ and $\log g_{\odot}$ is possible 
\citep[][]{melendez09:twins,ramirez09}.

Ten stars were identified as most closely resembling the Sun and are
given in Table \ref{t:twins}, including HIP56948, the best solar twin
currently known \citep[][]{melendez07,takeda09}. A crucial requirement
for these stars is to have accurate and homogeneous photometry in
order to derive reliable $\teff$ via IRFM. While this is possible in
the infrared because of 2MASS\footnote{In fact, the other well known
  solar twin 18 Sco \citep{demello97} has saturated 2MASS colours.},
optical photometry is also important to properly recover the
bolometric flux where these stars emit most of their
energy. Johnson-Cousins photometry would be the ideal choice, but
unfortunately is not available for all these targets. To overcome this
limitation, in the optical we used the Tycho2 $B_TV_T$ system which
uniformly and precisely covers the entire sky in the magnitude range
of our interest \citep[][]{hog2000}. Notice that we did not
transform $B_TV_T$ into $BV$ but instead implemented our IRFM code
to work directly on the Tycho2 system. Also, as discussed in Section
\ref{proscons} the absence of $(RI)_C$ photometry is not relevant for
stars hotter than $5000$\,K. All twins are closer than $72$~pc, where
reddening is expected to be zero or negligible: nearly all of them have
Str\"omgren photometry (Mel{\'e}ndez et al.~in prep.) and the 
\citet{schuster89} reddening calibration confirms indeed such a conclusion. 

\begin{table*}
\centering
\caption{Tycho2 and 2MASS photometry for our solar twins sample,
  together with the spectroscopic parameters and the effective
  temperatures determined via IRFM. For the latter, the errors are
  those arising from the photometry alone, not including the $15$\,K
  uncertainty in the zero point of our temperature scale. All twins
  have ``A'' quality flag and Read 1 mode in all 2MASS bands.}
\label{t:twins}
\begin{tabular}{ccccccccccccccc}
\hline\hline
HIP & $B_T$ & $\sigma_B$ & $V_T$ & $\sigma_V$ & $J$ & $\sigma_J$ & $H$ & $\sigma_H$ & $K_S$ & $\sigma_K$ & $T_{\rm{eff}}^{\rm{spec}}$ & $\logg$ & $\feh$ & $T_{\rm{eff}}^{\rm{IRFM}}$\,(K) \\    
 & & & & & & & & & & & $\pm20$\,K & $\pm 0.04$~dex& $\pm0.022$~dex &  \\
\hline
30502  &   9.483  & 0.019 &  8.706 & 0.013 & 7.474 & 0.029 &  7.139 & 0.029 &  7.069 & 0.024 & 5745 & 4.47 &  $-0.01$ & $5760 \pm 28$  \\
36512  &   8.498  & 0.015 &  7.786 & 0.011 & 6.517 & 0.020 &  6.213 & 0.027 &  6.154 & 0.024 & 5755 & 4.53 &  $-0.08$ & $5763 \pm 26$  \\
41317  &   8.613  & 0.015 &  7.868 & 0.010 & 6.610 & 0.023 &  6.289 & 0.038 &  6.206 & 0.024 & 5740 & 4.49 &  $-0.02$ & $5739 \pm 27$  \\
44935  &   9.522  & 0.021 &  8.783 & 0.015 & 7.548 & 0.019 &  7.260 & 0.034 &  7.171 & 0.024 & 5800 & 4.41 &   $0.07$ & $5803 \pm 30$  \\
44997  &   9.122  & 0.017 &  8.378 & 0.012 & 7.107 & 0.021 &  6.888 & 0.051 &  6.764 & 0.026 & 5790 & 4.52 &   $0.03$ & $5791 \pm 30$  \\
55409  &   8.793  & 0.017 &  8.066 & 0.011 & 6.811 & 0.019 &  6.493 & 0.042 &  6.419 & 0.021 & 5760 & 4.52 &  $-0.01$ & $5758 \pm 26$  \\
56948  &   9.462  & 0.017 &  8.748 & 0.012 & 7.477 & 0.019 &  7.202 & 0.026 &  7.158 & 0.018 & 5782 & 4.38 &   $0.01$ & $5801 \pm 25$  \\
64713  &  10.048  & 0.029 &  9.280 & 0.021 & 8.086 & 0.018 &  7.771 & 0.026 &  7.707 & 0.034 & 5815 & 4.52 &  $-0.01$ & $5853 \pm 36$  \\
77883  &   9.532  & 0.023 &  8.820 & 0.018 & 7.476 & 0.021 &  7.176 & 0.038 &  7.125 & 0.034 & 5695 & 4.39 &   $0.04$ & $5660 \pm 35$  \\
89650  &   9.708  & 0.023 &  8.996 & 0.017 & 7.781 & 0.029 &  7.506 & 0.034 &  7.431 & 0.033 & 5855 & 4.48 &   $0.02$ & $5864 \pm 35$  \\
\hline
\end{tabular}
\end{table*}

\subsection{A finely tuned absolute calibration}\label{tuning}

As for the Johnson-Cousins system, we based the absolute calibration
of the Tycho2 system on Vega \citep[][]{bohlin04a,bohlin07}, adopting
the $B_TV_T$ zero points of \citet{apellaniz07} and the corresponding
filter transmission curves of \citet{bessell2000}. 
\begin{figure*}
\centering
\includegraphics[scale=0.6]{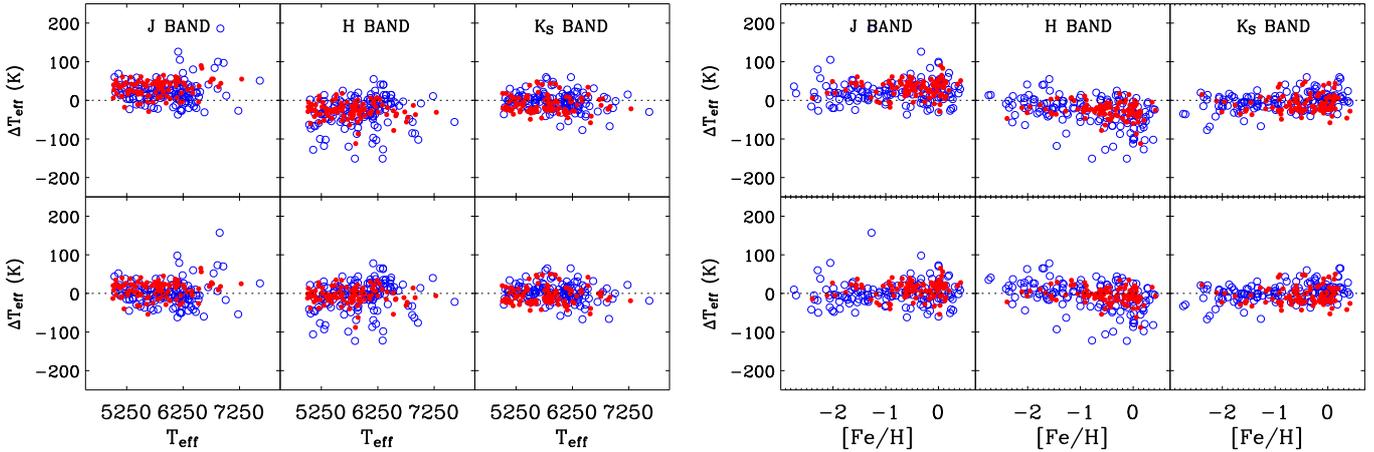}
\caption{Top panels: difference between $\teff$ and the effective
  temperature determined in each infrared band before tuning the
  absolute calibration. Full circles are stars with quality flag
  ``A'', Read 1 mode and total 2MASS photometric errors $< 0.07$
  mag.~while open circles are for all other stars. Bottom panels: as
  in the top panels, but with the adjusted absolute calibration. $H$
  band photometry has usually slightly higher error than $J$ and $K_S$
  and the final temperature is the weighted average of that obtained
  in each band.}
\label{f:off}
\end{figure*}

The first instance, we determined $\teff$ via IRFM for each of the
twins in Table \ref{t:twins}: their average effective temperature
turned out to be 5782\,K, remarkably close to $T_{\rm{eff},\odot}$,
thus confirming the high accuracy achieved using the HST and 2MASS
absolute calibration. Based on Monte Carlo simulations with the
photometric errors in Table \ref{t:twins}, the uncertainty in $\teff$
determined via IRFM is of order $30$\,K for single stars. Imposing the
mean effective temperature of all solar twins to equal
$T_{\rm{eff},\odot}$ we estimate the uncertainty on the zero point of
our temperature scale to be 15\,K based on a bootstrap procedure with
one million re-samples. At the same time, for HIP56948 we also recover
$T_{\rm{eff},\odot}$ within $1 \sigma$.  

Though the solar twins test confirms the global reliability of the
adopted absolute calibration, for all stars in Section \ref{sampleJC}
having Tycho2 photometry and $\teff>5000$\,K we further required each
infrared band to return on average the same $\teff$ as the others
(Figure \ref{f:off}). By imposing such a consistency we improve upon
small systematic trends which could arise when determining effective
temperatures in stars with $\teff$ and $\feh$ very different from our
solar twins. This led to a decrease of the absolute calibration by
$1.6$\% in the $J$ band and an increase by $1.5$ and $0.3$\% in the $H$ and
$K_S$ bands, respectively (see also Appendix A). 
In terms of synthetic magnitudes these
differences make $H$ and $K_S$ redder by 0.016 and 0.003 and $J$ bluer
by 0.017, thus removing almost entirely the infrared colour offsets
found by \citet{casagrande06} when comparing observed and synthetic
photometry. We cannot entirely rule out whether these systematic
differences arise from the adopted synthetic library or the absolute
calibration, but since the IRFM depends only marginally on model
atmospheres and the infrared spectral region is relatively easy to
model, we are strongly in favour of the second possibility. From a
pragmatic point of view, this improves the consistency in determining
$\teff$. Also, such changes are within the 2MASS quoted errors and for
the $K_S$ band we remark the agreement with the $0.2$\% increase found
by \citet{rieke08} and discussed in Section \ref{proscons}.  
As expected, stars with the best 2MASS pedigree also return better
agreement in all bands (full circles in Figure \ref{f:off}). We have
also checked that the increasing scatter in Figure \ref{f:off} is
primarily due to photometric errors. We recall that \citet{rieke08}
found a 2\% offset between Read 1 and Read 2 mode in
2MASS\footnote{This mode indicates which readout is used to derive
  photometry
  http://www.ipac.caltech.edu/2mass/releases/allsky/doc/sec3\_1b.html},
though they were not able to derive a universal correction for this
effect. All our solar twins have Read 1 mode and the absence of a
universal correction suggests that while Read mode 2 can decrease the
precision of $\teff$ the overall accuracy of our calibration remains
valid. 

With the fine-tuning discussed above, the median (mean) effective
temperature of our solar twins is 5777 (5779)\,K. 
Restricting only to the twins having $T_{\rm{eff},\odot}$ within 
the observational errors, still confirm such conclusion.
As a further independent test, we applied our IRFM to the list of solar analogs
used by \citet{rieke08} and determined their median (mean) $\teff$ to
be 5791 (5786)\,K, thus confirming the reliability of the zero point
of our temperature scale, which has an uncertainty of 15\,K. Such a
value implies possible systematics in the absolute calibration at the
1\% level. The systematic error in recovering the bolometric
luminosity is however smaller since infrared fluxes enter twice in
$R_{obs}$, thus partly compensating their uncertainty.   

The corrections in the infrared absolute calibration discussed here 
have been used also in determining $\teff$ for stars in Section
\ref{sampleJC}. Since for those stars we are using Johnson-Cousins
photometry, there could still be small differences arising from the
absolute calibration in the optical: for stars in common a mean
systematic of 8\,K in $\teff$ and $0.15$\% in bolometric flux was
found and corrected. 

\section{Validating the proposed temperature scale}\label{valid}

The IRFM determines $\teff$ in an almost model independent way,
primarily recovering the bolometric flux $\mathcal{F}_{Bol}\rm(Earth)$
of the star under investigation. From the basic definition linking
those two quantities the stellar angular diameter $\theta_{\rm{IRFM}}$
can be obtained self-consistently and this was actually one of the driving
reasons for developing the technique \citep{blackwell77}. In what
follows, we use this information to further validate our results.  

\subsection{Interferometric angular diameters}\label{inter}

An independent test of accuracy for the zero point of our effective 
temperature scale involves the comparison with the angular diameters
measured using interferometric techniques (corrected for
limb-darkening, hereafter denoted by $\theta_{\rm{LD}}$).
In our case, angular diameters are a natural 
consequence of the $\teff$ determination procedure and for each star
the $\teff,\mathcal{F}_{Bol},\theta_{\rm{IRFM}}$ values are self-consistent,
i.e., they represent a unique solution for a given set of input
data. We also prefer to compare angular diameters directly
(i.e.~$\theta_{\rm{IRFM}}$ vs.~$\theta_{\rm{LD}}$) since the effective
temperatures reported in various interferometric works would be more
heterogeneous because of the adopted bolometric corrections.
\begin{figure}
\includegraphics[scale=0.61]{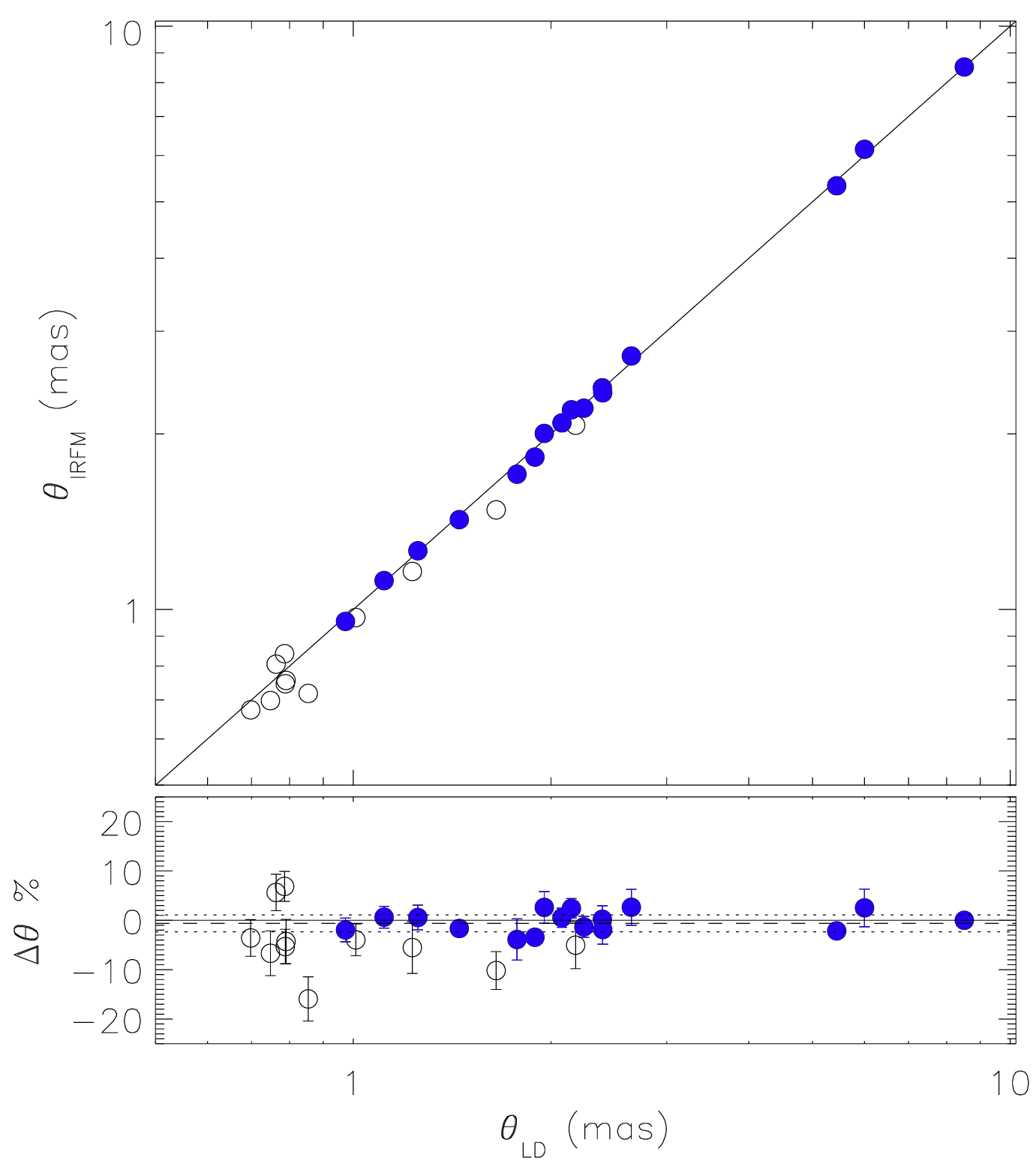}
\includegraphics[scale=0.61]{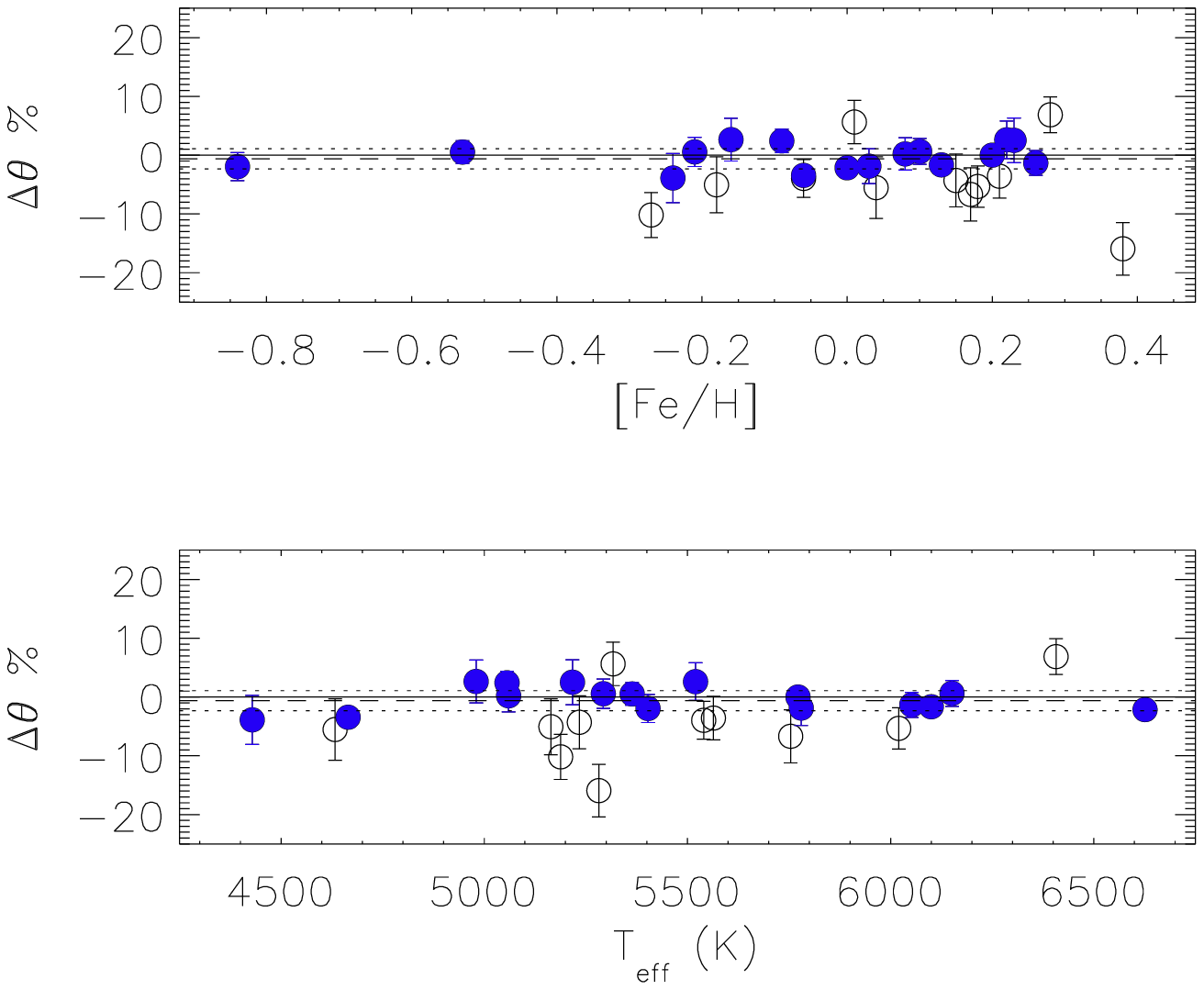}
\caption{Top two panels: Comparison of angular diameters measured
  interferometrically ($\theta_\mathrm{LD}$) and via our IRFM
  photometric calibrations ($\theta_\mathrm{IRFM}$). Full symbols
  represent stars that have $\theta_\mathrm{LD}$ measured with
  accuracy better than 2\,\%. Bottom two panels: Difference (in \%)
  between $\theta_\mathrm{LD}$ and $\theta_\mathrm{IRFM}$ as a
  function of stellar parameters. Solid lines represent 1-to-1
  correspondence, dashed and dotted lines are the average difference
  and 1-$\sigma$ error for the full data point, respectively.}
\label{f:ang_diam}
\end{figure}

Given the difficulties involved in the measurement of the small
angular diameters of dwarfs and subgiants (even the nearest ones have
angular diameters below 10 milli-arcseconds), only a
relatively small group of such stars has been observed to date for
that purpose (see also Appendix A for a discussion of the angular 
diameters used by Alonso et al.~1994a). 
We performed a literature search for interferometrically
determined angular diameters with precision better than 5\,\% (which 
corresponds to an accuracy of 2.5\% in effective temperatures, roughly
$150$\,K at solar temperature, assuming no error in the bolometric flux)
and found data for 28 stars, 16 of which have $\theta_{\rm{LD}}$ measured
to better than 2\,\% (Table~\ref{t:ang_diam}). The efforts made by the
interferometry community in the last few years are commendable given
that the number of stars with reliable $\theta_\mathrm{LD}$ has nearly
doubled since 2005 \citep[cf.][]{ramirez05a}. 

Unfortunately, all dwarfs and subgiants with reliable
$\theta_\mathrm{LD}$ are brighter than $V\simeq6$, implying
infrared magnitudes $\lesssim 5$ where 2MASS photometry has large
observational errors and starts to
saturate\footnote{\tiny{www.ipac.caltech.edu/2mass/releases/allsky/doc/sec2\_2.html\#pscphotprop}}. Therefore
we cannot apply our IRFM directly on them to get
$\theta_\mathrm{IRFM}$. Instead, we adopt an indirect approach using
the photometric $\teff$:colour and $\mathcal{F}_{Bol}$:colour relations 
presented in
Section \ref{colte}. Using the photometry of our sample stars (i.e.~those
used in the construction of the calibrations and therefore with
$\teff$ directly determined via IRFM), we checked that the zero point
of our $\teff$ and $\mathcal{F}_{Bol}$ scales is correctly reproduced
by the calibration formulae presented in
Section \ref{colte}, independently of the apparent magnitudes of the
stars. Also, for the two stars having HST 
spectrophotometry (next Section) we checked that our calibration 
formulae reproduce nearly the same results as directly applying the 
IRFM. 
We were careful about propagating all possible sources of
random error such as uncertainties in the input photometry,
metallicity, and the reliability of the colour calibrations, as
quantified by the standard deviation of each polynomial fit (Tables
\ref{tab:ct} and \ref{tab:cf}). For most of the stars with reliable
$\theta_\mathrm{LD}$ (i.e.~better than 2\%), only $BV$ photometry was
available, while for the remaining $BV(RI)_C$ was used.
Metallicities were adopted from the updated version of the \citet{cayrel01}
$\feh$ catalog by Mel\'endez~(in prep.), which nearly
triples the number of entries in the original catalog. 

The comparison of the angular diameters measured interferometrically with
those derived using our IRFM colour calibrations is shown in
Fig.~\ref{f:ang_diam} (see also Table \ref{tab:theta}). Stars that have
$\theta_\mathrm{LD}$ determined with accuracy better than 2\,\% are
shown with full symbols. Using only the latter, the average difference
in angular diameter (IRFM-LD) is $-0.62\pm1.70$\,\% which corresponds 
to a zero point difference in the effective temperature scale of only
$+18\pm50$\,K at solar temperature. This is also in agreement with the
uncertainty on the zero point of our temperature scale discussed in
Section \ref{tuning}. No obvious trends are seen with
$\feh$ (from about $-0.8$ to $+0.3$) or $\teff$ (from 4400 to
6600\,K). Note, however, that if we exclude the two coolest stars
(from the group of those having errors smaller than 2\,\%),
a small trend is seen with $\teff$. The trend --if real-- appears more clearly 
for early type stars, with $\theta_\mathrm{IRFM}$ being underestimated (and 
therefore the IRFM effective temperatures overestimated) with respect to the 
interferometric measurements. Interferometry resorts on 1D model atmospheres 
to correct from the measured uniform-disk angular diameter to the physical 
limb-darkened disk to which we compare with. Interestingly, 3D models predict 
less center-to-limb variation than 1D models as moving from K to F type stars 
\citep{allende02,bigot06}. Reduced limb-darkening corrections imply smaller 
$\theta_{\rm{LD}}$: the trend discussed above qualitatively fit into this 
picture. How well our result agrees quantitatively with this picture we leave 
to future studies. 

Interestingly, \citet{ramirez05a} made a similar comparison of 
angular diameters and also found good agreement with their IRFM $\teff$ 
scale, which is, however, systematically cooler (by $\simeq100$\,K) than the 
present one for $\feh \gtrsim -2$ 
\citep[see also][]{casagrande08:uppsala}. 
We compared the stars with 
angular diameters in common between table 4 of 
\citet[][RM05]{ramirez05a} and the present study (C09, Table 
\ref{tab:theta}) and found an average difference (C09-RM05) of 
$0.1\pm2.2$\,\% in angular diameters, $3.0\pm3.0$\,\% in 
bolometric fluxes and $40 \pm 37$~K in $\teff$. Given the large 
scatter, these numbers are still consistent with the mean 
differences in $\teff$ and $\mathcal{F}_{Bol}$ from these two 
studies (Section \ref{comparisonRM}), however, we would expect our diameters 
to be roughly
smaller by 3\,\%, our fluxes brighter by 1\,\% and our $\teff$ 
hotter by 100~K \citep[see also][]{casagrande06}. While $\mathcal{F}_{Bol}$ 
and $\teff$ compensate to give almost exactly the same angular 
diameters, the $40$~K offset might be more representative of the 
difference with the TCS magnitudes used in \citet{ramirez05a} (see 
the discussion on the small zero point differences to convert 
2MASS into TCS presented in Section \ref{comparisonRM}).
To gauge further insights, we redetermined the temperatures used 
by \citet{ramirez05a} using their colour calibrations for the same 
$BV(RI)_C$ input data we adopted in this section and found 
$\Delta \teff = 72 \pm 52$~K. In addition, we 
adopted our bolometric fluxes lowered by 1\,\%, which corresponds 
to the average difference we find for our complete sample. In this 
case the difference in angular diameters sets to $-2.4\pm2.1$\,\%, 
much closer to the expected $-3$\,\%, offsetting the 
\citet{ramirez05a} scale with respect to interferometric 
measurements. Since the present work represents an improvement 
over \citet{ramirez05a}, in particular the fact that the 
$\teff,\mathcal{F}_{Bol},\theta_\mathrm{IRFM}$ values are a 
self-consistent and unique solution to each problem star, and given
that the number of comparison stars has doubled since 2005 (note also
that the $\theta_\mathrm{LD}$ values of some stars have been 
re-determined), it is likely that the good agreement found by 
\citet{ramirez05a} was due to a conspiracy of photometric errors 
which propagated to both $\teff$ and $\mathcal{F}_{Bol}$ 
determinations and low number statistics. More measurements of 
stellar angular diameters via interferometry are clearly necessary, and 
therefore highly encouraged, to better constrain indirectly 
determined effective temperature scales. However, as this exercise has shown, 
many critical ingredients enter in the comparison with angular 
diameters. In particular bolometric corrections and effective 
temperatures should be determined as self-consistently as possible, 
also avoiding transformation between photometric systems. It gives us 
confidence that the zero point uncertainty from solar twins, angular 
diameters and HST spectrophotometry (next Section) returns in all cases 
independent and very consistent results.

While the angular diameter comparison does not extend below $\feh\simeq-1.0$, 
leaving our results for halo stars ``un-tested'' in this context, in the next 
Section we use HST spectrophotometry to gauge further insight on the
topic. 

\begin{table}
\caption{Stars with measured interferometric angular diameters.}
\label{tab:theta}
\begin{tabular}{rcccrc} \hline\hline
HD & $\theta_\mathrm{LD}$ & Ref.$^{\mathrm{a}}$ & $\teff^{\rm{IRFM}}$ & $\feh$ & $\theta_\mathrm{IRFM}$ \\  
   & mas & & K & dex & mas \\ \hline

  3651 & $0.790\pm0.027 $ & 1 & 5234 & $ 0.15$ & $0.756\pm0.022$ \\
  6582 & $0.973\pm0.009 $ & 2 & 5403 & $-0.84$ & $0.954\pm0.021$ \\
  9826 & $1.114\pm0.009 $ & 1 & 6151 & $ 0.10$ & $1.121\pm0.023$ \\
 10700 & $2.078\pm0.031 $ & 3 & 5364 & $-0.53$ & $2.089\pm0.026$ \\
 10780 & $0.763\pm0.021 $ & 2 & 5317 & $ 0.01$ & $0.806\pm0.022$ \\
 19994 & $0.788\pm0.026 $ & 1 & 6020 & $ 0.18$ & $0.746\pm0.009$ \\
 22049 & $2.148\pm0.029 $ & 3 & 5056 & $-0.09$ & $2.200\pm0.032$ \\
 23249 & $2.394\pm0.029 $ & 3 & 5060 & $ 0.08$ & $2.399\pm0.059$ \\
 26965 & $1.650\pm0.060 $ & 3 & 5188 & $-0.27$ & $1.482\pm0.018$ \\
 61421 & $5.443\pm0.030 $ & 3 & 6626 & $ 0.00$ & $5.326\pm0.068$ \\
 75732 & $0.854\pm0.024 $ & 1 & 5282 & $ 0.38$ & $0.718\pm0.025$ \\
102870 & $1.450\pm0.018 $ & 4 & 6100 & $ 0.13$ & $1.426\pm0.014$ \\
117176 & $1.009\pm0.024 $ & 1 & 5540 & $-0.06$ & $0.969\pm0.021$ \\
120136 & $0.786\pm0.016 $ & 1 & 6407 & $ 0.28$ & $0.840\pm0.019$ \\
121370 & $2.244\pm0.019 $ & 3 & 6052 & $ 0.26$ & $2.214\pm0.043$ \\
128620 & $8.511\pm0.020 $ & 3 & 5772 & $ 0.20$ & $8.511\pm0.079$ \\
128621 & $6.000\pm0.021 $ & 5 & 5217 & $ 0.23$ & $6.151\pm0.234$ \\
131977 & $1.230\pm0.030 $ & 3 & 4633 & $ 0.04$ & $1.162\pm0.054$ \\
150680 & $2.397\pm0.044 $ & 3 & 5780 & $ 0.03$ & $2.352\pm0.055$ \\
161797 & $1.953\pm0.039 $ & 3 & 5520 & $ 0.22$ & $2.004\pm0.050$ \\
185144 & $1.254\pm0.012 $ & 2 & 5293 & $-0.21$ & $1.261\pm0.029$ \\
188512 & $2.180\pm0.090 $ & 6 & 5164 & $-0.18$ & $2.070\pm0.049$ \\
190360 & $0.698\pm0.019 $ & 1 & 5564 & $ 0.21$ & $0.673\pm0.017$ \\
198149 & $2.650\pm0.040 $ & 3 & 4980 & $-0.16$ & $2.720\pm0.090$ \\
201091 & $1.775\pm0.013 $ & 3 & 4429 & $-0.24$ & $1.706\pm0.070$ \\
209100 & $1.890\pm0.020 $ & 3 & 4665 & $-0.06$ & $1.825\pm0.021$ \\
217014 & $0.748\pm0.027 $ & 1 & 5754 & $ 0.17$ & $0.698\pm0.019$ \\
			    
\hline
\end{tabular}
\begin{list}{}{}
\item[$^{\mathrm{a}}$] 1.-- \citet{baines08}, 2.-- \citet{boyajian08},
  3.-- \citet{kervella08} (weighted average if more than one
  measurement was available), 4.-- \citet{north09}, 5.-- \citet{bigot06},
  6.-- \citet{nordgren99}.
\end{list}
\label{t:ang_diam}
\end{table}

\subsection{HST spectrophotometry}

For each star, we obtain a synthetic spectrum tailored at the
effective temperature determined via IRFM (Section
\ref{proscons}). Since the angular diameter is determined, each
synthetic spectrum is absolutely calibrated (i.e.\, in units of
$\rm{erg\,cm^{-2}\,s^{-1}}\,\AA^{-1}$), and can be used to further
test our results. In fact, from F- to early K-type stars, all
continuum characteristics approximately longward of the Paschen
discontinuity depend almost exclusively on the effective temperature,
relatively unaffected by spectral lines and NLTE effects as well as
from the treatment of convection.   
\begin{figure*}
\centering
\includegraphics[scale=0.7]{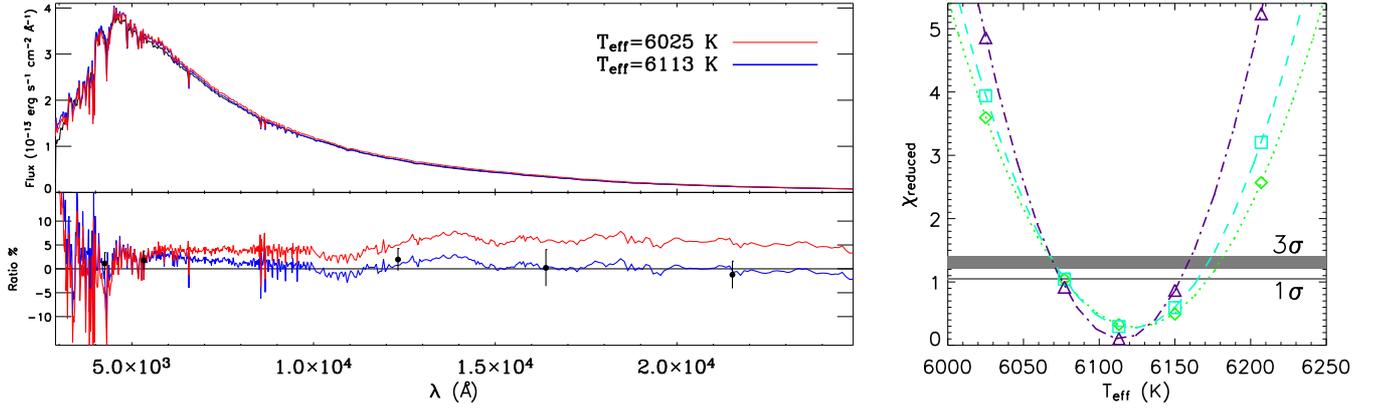}
\caption{Left upper panel: comparison between the observed HD209458 CALSPEC spectrum (black line) and the synthetic spectra derived for two different $\teff$, using our preferred absolute calibration (blue line) and increasing the infrared absolute calibration by 5\% (red line). Left lower panel: ratio of synthetic to observed spectra. Full circles are the ratio between the fluxes obtained once the Vega calibration is used with the observed magnitudes and the fluxes obtained directly from the convolution of the CALSPEC spectrum with the appropriate filter transmission curve. Error bars take into account uncertainty in the Vega calibration and zero points, as well as in the observed magnitudes. Right panel: reduced $\chi^{2}$ for various $\teff$ solutions corresponding to different adopted absolute calibrations. Our choice (Section \ref{tuning}) always lies very close to the minima obtained fitting a parabola to the data (lines of different style). Different symbols correspond to cut longward of $0.66\,\mu$m (diamonds), $0.82\,\mu$m (squares) and $1.46\,\mu$m (triangles) as explained in the text. The sigma levels have been computed using the incomplete gamma function for the number of degrees of freedom longward of our cuts.}
\label{f:hd}
\end{figure*}

The
CALSPEC\footnote{http://www.stsci.edu/hst/observatory/cdbs/calspec.html
as of January 2009.} library contains composite stellar spectra measured by the STIS ($0.3-1.0\,\mu\rm{m}$) and NICMOS ($1.0-2.5\,\mu\rm{m}$) instruments on board of the HST and used as fundamental flux standard. Free of any atmospheric contamination the HST thus provides the best possible spectrophotometry to date, with $1-2$\% accuracy, extending from the far-UV to the near infrared. The absolute flux calibration is tied to the three hot, pure hydrogen white dwarfs, which constitute the HST primary calibrators, normalized to the absolute flux of Vega at 5556\,\AA\,\citep{bohlin07}. Thus, except for the normalization at 5556\,\AA\, the absolute fluxes measured by STIS and NICMOS are entirely independent on possible issues regarding Vega's absolute calibration in the infrared and offer an alternative approach to the 2MASS calibration underlying our temperature scale.

Two of the CALSPEC targets are late-type main-sequence dwarfs for
which accurate photometry, $\logg$ and $\feh$ are available: the
exoplanet host star HD209458 \citep[e.g.][]{charbonneau00} and the
fundamental SDSS standard BD $+17\,4708$
\citep[e.g.][]{fukugita96,smith02}. For each of these targets we
computed $\teff$ and derived the corresponding physical flux using the
absolute calibration presented in Section \ref{tuning}. For
comparison, we also determined the effective temperatures and the
corresponding fluxes when changing our adopted infrared absolute
calibration by different amounts up to $\pm 5$\%, which roughly correspond 
to $\mp 100$\,K in $\teff$.
The agreement was quantified using $\chi^{2}$ statistics between the observed ($\mathcal{F}$) and synthetic ($\tilde{\mathcal{F}}$) spectra at various $\teff$
\begin{equation}
\chi^2 = \sum_{\lambda}\frac{\left(\mathcal{F}_{\lambda}-\tilde{\mathcal{F}}_{\lambda}\right)^2}{\sigma_{\lambda}^2}
\end{equation}
where $\sigma_{\lambda}^2$ is the squared sum of the CALSPEC and our random errors, arising primarily from the photometry and to minor extent $\feh$ and $\logg$. Angular diameters are needed to scale synthetic spectra into physical units: typical $1$\% internal accuracy in $\theta_{\rm{IRFM}}$ implies $2$\% errors in the derived flux. We decided to use random errors only because the purpose of the test is exactly to verify the range of values allowed once the zero point of the temperature scale is assumed. 

Also, the tuning of the absolute calibration in the infrared affects the final $\teff$ but it does not modify in any manner the shape of the synthetic spectrum, which entirely depends on the \citet{castelli04} grid interpolated at the proper $\teff$, $\logg$ and $\feh$. Notice that we are not searching for the synthetic spectrum which best matches the observation, rather we want to test the effective temperature we derive: while adjustments to $\feh$ and $\logg$ could improve the agreement in the blue and visible part, the continuum characteristics are more sensitive to $\teff$.
\begin{figure*}
\centering
\includegraphics[scale=0.7]{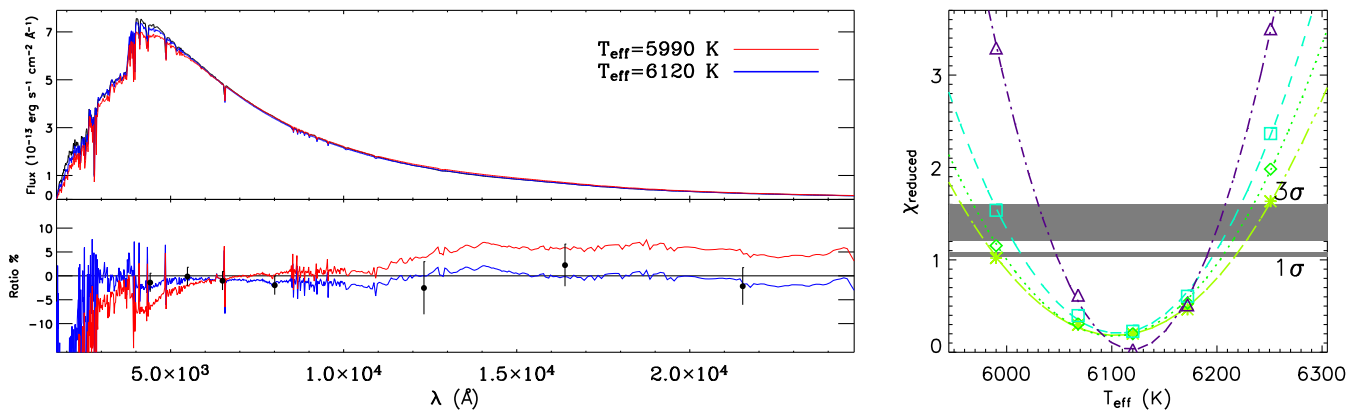}
\caption{Same as in figure \ref{f:hd} for BD $+17\,4708$. The synthetic spectra have been reddened by $E(B-V)=0.01$. Different symbols in the right panel correspond to cut longward of $0.50\,\mu$m (asterisks) $0.66\,\mu$m (diamonds), $0.82\,\mu$m (squares) and $1.46\,\mu$m (triangles). The maximum wavelength used for computing the reduced $\chi^2$ has been $2\,\mu$m to avoid possible contribution from the cool companion.}
\label{f:bd}
\end{figure*}

\subsubsection{HD209458}

For this target we adopted the spectroscopic $\feh=0.03 \pm 0.02$ and $\logg=4.50 \pm 0.04$ measured from the high precision HARPS GTO sample \citep{sousa08} and used Tycho2 and 2MASS photometry. We obtain
$\teff=6113 \pm 49$\,K, $\mathcal{F}_{Bol}=(2.335 \pm 0.025) \times 10^{-8} \,\rm{erg\,cm^{-2}\,s^{-1}}$ and $\theta=0.224 \pm 0.004$\,mas 
including both random and systematic errors. The latter result is in
good agreement with the angular diameters $0.215 \pm 0.009$\,mas
obtained using the new {\it Hipparcos} parallaxes \citep{vanLeeuwen07}
to convert the linear radius measured from exoplanet transit
photometry with HST \citep{brown01}. Notice that $\sim 100$\,K cooler
effective temperatures would imply values of $\theta$ larger by $\sim
3.5$\% in the IRFM. 

The comparison between the observed and synthetic spectra at two different $\teff$ is shown in Figure \ref{f:hd}: while they both succeed to capture the main observed features, the continuum of the cooler model is clearly off from the observation. We quantify the agreement between the HST spectrophotometry and the models at various $\teff$ applying $\chi^2$ statistics longward of the $\rm{H}\alpha$ line ($0.66\,\mu$m), the Paschen ($0.82\,\mu$m) and the Brackett ($1.46\,\mu$m) discontinuity. These cuts define the beginning of the continuum in a somewhat arbitrary manner, but they all return consistent results thus ensuring that our conclusion is not affected by their choice. 
The reduced $\chi^{2}$ is lower than $1$ in a roughly $\pm 40$\,K interval effectively centered on our preferred solution. While reduced $\chi^{2} < 1$ tells that the size of the errors is still too large to clearly favour a solution within that range, the large number of points used in the test sets low $1 \sigma$ and $3 \sigma$ levels, clearly ruling out solutions different by $\pm100$\,K.

\subsubsection{BD $+17\,4708$}

This star is the only subdwarf with well measured absolute flux, thus making it an important benchmark for testing the temperature scale in the metal-poor regime. We adopt the spectroscopic parameters $\feh = -1.74 \pm 0.09$, $\aFe=0.4$ and $\logg =3.87 \pm 0.08$ from \citet{ramirez06} who also derived $\teff = 6141 \pm 50$\,K, $\mathcal{F}_{Bol}=(4.89 \pm 0.10) \times 10^{-9} \,\rm{erg\,cm^{-2}\,s^{-1}}$ and $\theta = 0.1016 \pm 0.0023$\,mas.
We corrected for reddening $E(B-V)=0.01$ the optical (Table \ref{newUBVRI}) and infrared (2MASS) magnitudes, obtaining $\teff = 6120 \pm 112$\,K, $\mathcal{F}_{Bol}=(4.80 \pm 0.04) \times 10^{-9} \,\rm{erg\,cm^{-2}\,s^{-1}}$ and $\theta=0.101 \pm 0.003$ all in excellent agreement with the aforementioned analysis. Radial velocities show modulation consistent with the presence of a low mass companion which could influence infrared photometry \citep{latham88}. The flags associated with 2MASS indicate excellent quality and no artifact nor contamination in any band, pointing toward a negligible effect, if any. Nonetheless, since the percent contribution of a cool companion increases with increasing wavelength, as safety rule we decided not to use $K_S$ in the IRFM though it would change the resulting $\teff$ by only $12$\,K. 
For our preferred $\teff=6120$\,K, shortward of $2\,\mu$m there is an outstanding agreement with the CALSPEC observed spectrum, meaning that the solution found represents well the observation at all wavelengths. A moderate increase in the observed with respect to the synthetic flux seems to appear longward of $2\,\mu$m, which could be the signature of the cooler companion. On the contrary, cooler solutions overestimate the flux throughout the entire continuum.

Because of the metal-poor nature of this star, the continuum shows up already at bluer wavelengths. We compute the reduced $\chi^2$ in different intervals, starting longward of $0.50\,\mu$m: as for the previous star, our solution substantially correspond to the minima of all parabolae, independently of the cut adopted. The random errors associated with this star are larger than in the case of HD209458, giving shallower minima and thus making it more difficult to discriminate between different solutions. However, differences up to $\pm 100$\,K are clearly disfavoured (Figure \ref{f:bd}).

Summarizing, CALSPEC data support our temperature scale which provide the best match to the observed spectrophotometry, in both metal-rich and -poor regimes. While differences larger than $\pm 40$\,K are ruled out for HD209458, the observational errors for the metal-poor star allow bigger uncertainties. Nonetheless, we have determined the fundamental parameters of both stars with the same procedure and in both cases our solutions are located at the minimum $\chi^2$: we regard such a result as a further indication that our $\teff$ scale is well calibrated over a wide metallicity range. 

\section{The new effective temperature scale}

Our results should be compared with effective temperatures determined employing different methods. First, we focus on large studies which have targeted solar neighbourhood stars, where the vast number of objects imposes the use of fast and efficient techniques, relying on fitting the observed photometry or spectra to their synthetic counterpart. An extensive comparison between the effective temperatures determined from high resolution spectroscopy of solar neighbourhood stars and a version of the IRFM similar to that adopted here has been already carried out in \citet{sousa08}. 
For metal-poor stars we restrict the comparison to purely spectroscopic effective temperatures; their validation will be crucial for ongoing and future studies of halo stars which are strongly affected by reddening and often lacking photometry. 

\subsection{Solar Neighbourhood stars}

\subsubsection{Valenti \& Fischer sample}

\begin{figure}
\includegraphics[width=9.0cm]{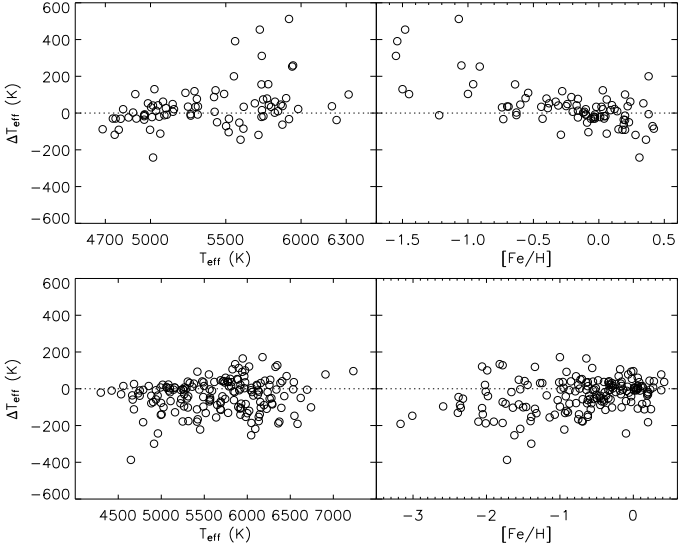}
\caption{Upper (lower) panels: comparison between the effective temperatures determined in this work and those obtained by \citet{valenti05} (Masana et al. 2006).$\Delta \teff$ are this $-$ other works in all panels.}
\label{f:vm}
\end{figure}

\citet{valenti05} have presented a uniform catalogue of stellar properties for 1040 nearby F,G and K stars which have been observed by the Keck, Lick and AAT planet search programs. Fitting the observed spectra with synthetic ones, they have obtained effective temperatures, surface gravities and abundances for every star. For 84 objects in common, there is no obvious dependence as a function of $\teff$, except for a drift appearing below $5000$\,K. However, when $\Delta \teff$ is plotted as function of metallicity the trend becomes clear, with very significant discrepancies at the lowest metallicities (Figure \ref{f:vm}).

\subsubsection{Masana et al. sample}

\citet{masana06} have derived stellar effective temperatures and bolometric corrections by fitting $V$ and 2MASS IR photometry. They calibrate their scale by requiring a set of 50 solar analogs drawn from \citet{cayrel96} to have on average the same temperature as the Sun. 

We have 176 stars in common: there is no obvious trend with effective temperatures, and for metallicities around solar there is an overall good agreement. This is not entirely unexpected considering that both studies have been calibrated to the Sun (though with different approaches): considering $\feh >-1$ the mean difference (IRFM $-$ Masana) is $\Delta \teff=-21 \pm 6$\,K ($\sigma=71$\,K).
However, when focusing on metal-poor stars $\feh < -1$ there is a significantly increasing scatter and a trend resulting in our $\teff$ being cooler up to $\sim 200$\,K at the lowest metallicities and with a mean difference of $-95 \pm 22$\,K ($\sigma=157$\,K).

\subsection{Metal-poor, halo stars}

\subsubsection{Temperatures from fits to hydrogen line profiles}\label{Balmer}

The wings of hydrogen lines are strongly sensitive to the effective
temperature of the star and only mildly dependent on the other stellar
parameters, other than being unaffected by reddening. Such approach is
particularly effective with metal-poor stars, given the lack of
severe line blending affecting the hydrogen lines. Thus, provided a
proper continuum normalization is applied, which can be non-trivial in
some cases \citep[e.g.][]{barklem02}, these lines can be used to determine 
$\teff$. Although
significant progress has been made in the last few years, the modeling
of hydrogen lines (e.g., the Balmer line profiles) is still quite
uncertain \citep{barklem00,barklem07}. Nonetheless, the relative
$\teff$ values derived in this manner can be very precise
\citep[e.g.][]{nissen07}. 

We remark that there is no such thing as one Balmer line $\teff$ scale, but 
instead each study depends upon the adopted prescriptions: LTE vs.~NLTE, 
broadening recipes, mixing-length parameter and even the details on how lines are fitted. Also, the 
thermal structure of the model atmosphere is crucial for the Balmer 
temperatures: as concerns 1D models, studies relying on OS- instead of 
ODF-model atmosphere determine hotter $\teff$ \citep{grupp04}. 

\begin{figure}
\includegraphics[width=9.0cm]{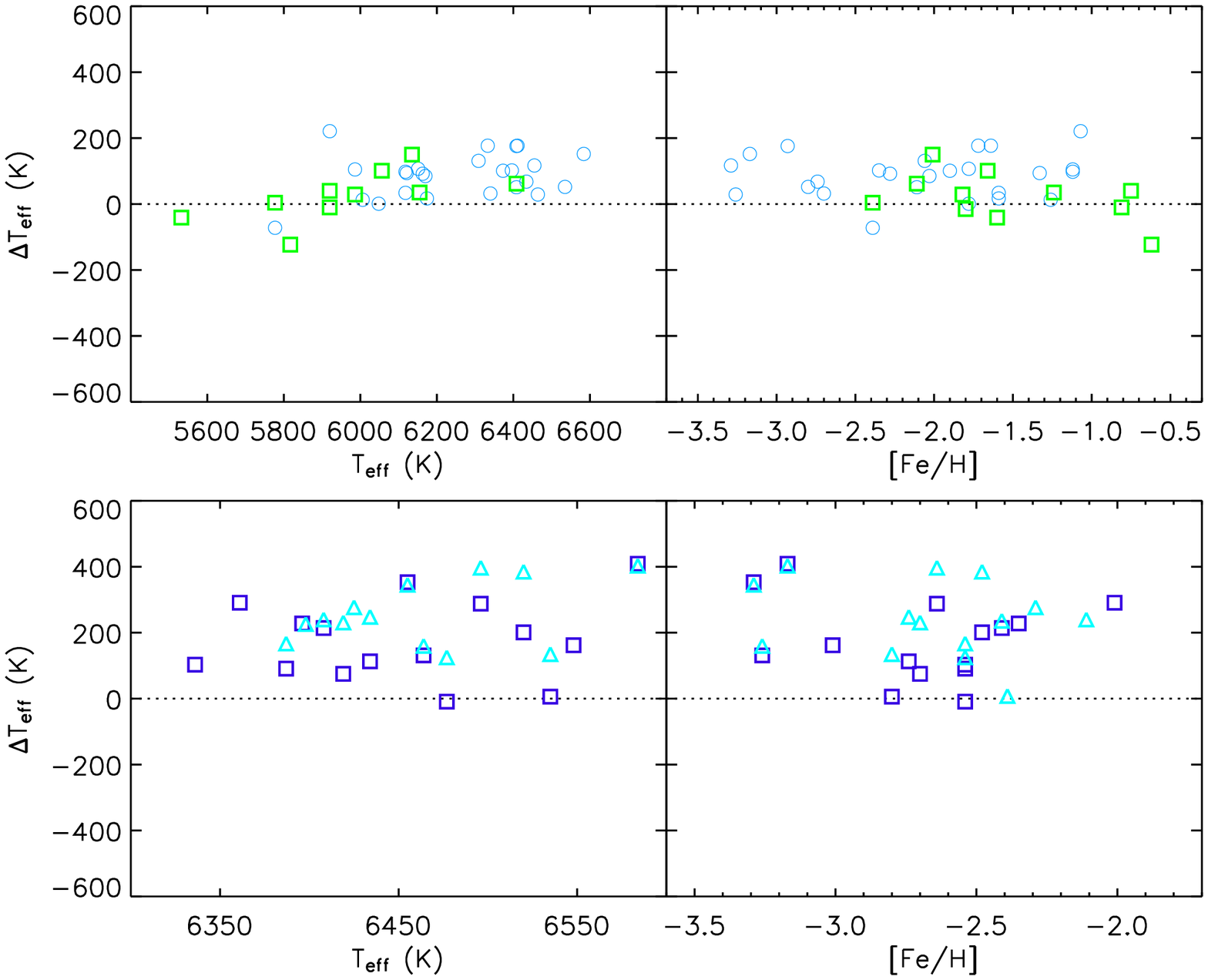}
\caption{Upper panels: comparison between the effective temperatures determined in this work and those obtained from the H$\beta$ \citep[][circles]{fabbian09} and H$\alpha$ plus H$\beta$ \citep[][squares]{bergemann} line profiles. Lower panel: comparison with respect to the excitation equilibrium temperatures determined by \citet{hosford09}. Two sets of data points are shown because \citet{hosford09} temperatures are sensitive to the uncertain $\logg$ values of metal-poor stars; squares (triangles) represents $\teff$ derived assuming the star to be on the main-sequence (sub-giant branch). $\Delta \teff$ are this $-$ other works in all panels.}
\label{f:fbh}
\end{figure}

Aware of the complexity of the picture, in the upper panels of Figure 
\ref{f:fbh} our IRFM effective temperatures are compared with those derived 
from fits to the Balmer lines in two different studies, which we regard as 
representative of the LTE and NLTE approach, respectively. 
Circles refer to the comparison with
\citet{fabbian09} who used the H$\beta$ lines. There is an obvious
offset, the IRFM returning $\teff$ hotter by $84 \pm 13$\,K ($\sigma =
66$\,K), but the small scatter between these two sets further
strengthen the conclusion that both techniques have high internal
precision. A similar conclusion holds also from the comparison with the 
effective temperatures reported in \citet[][,and references therein]{bergemann} 
who used both H$\alpha$ and H$\beta$ line profiles. In this case the 
difference (IRFM\,$-$\,H lines) is $21 \pm 23$\,K ($\sigma = 72$\,K) with a 
possible trend suggesting excellent agreement roughly below 6000\,K (one star, 
HD25329 with $\teff = 4785$\,K and $\Delta \teff=-15$\,K is not shown in the 
upper left panel of Figure \ref{f:fbh}). 

\subsubsection{Excitation equilibrium temperatures}

An important number of iron lines are present in the spectra of cool dwarfs, even the metal-poor ones. In an ideal case, the iron abundances determined from each of those lines should be consistent with each other. In practice, however, given an initial set of stellar parameters, the line-by-line abundances show trends with excitation potential (EP) and/or reduced equivalent width. By tuning the stellar parameters, these trends can be eliminated. The EP trend is particularly sensitive to $\teff$, given the strong dependence of the atomic level populations on temperature, and therefore $\teff$ determined by removing the abundance vs.~EP trend are often referred to as ``excitation equilibrium'' temperatures. Because of its nature, this method of $\teff$ determination is highly model-dependent. Not only it does require realistic model atmospheres and spectrum synthesis, but also accurate atomic data and, ideally, a non-LTE treatment of the line formation. The advantage of such method is that it is independent of interstellar reddening and can be applied to stars with uncertain or unavailable photometry.

Recently, \citet{hosford09} have determined LTE excitation equilibrium
temperatures for a sample of metal-poor stars. The difference found
between their temperatures and ours is illustrated in Figure
\ref{f:fbh} (HD140283 with $E(B-V)=0.000$, $\teff=5777$\,K and 
$\Delta \teff=8$\,K is not shown in the lower left panel). Because the 
excitation temperatures are somewhat sensitive to $\logg$ and surface gravities
of metal-poor stars are difficult to determine due to
uncertain/unavailable parallaxes, they provide two sets of $\teff$
values, one assuming the star to be on the main-sequence (MS) and
another one assuming the star to be on the subgiant branch (SGB). 
We remark that for HD140283 parallax and Balmer jump rule out the 
main-sequence stage; our fit (Mike Bessell) of the MILES fluxes using 
\citet{munari05} spectral library provide $\teff=5812/5875$\,K and 
$\logg=3.75$ for $E(B-V)=0.000/0.017$, respectively. 

The IRFM temperatures are significantly hotter than the excitation
temperatures by $177 \pm 33$\,K ($\sigma = 122$\,K) (for their MS
temperatures) and $240 \pm 32$\,K ($\sigma = 116$\,K) (SGB). In
particular, the large scatter suggests a decreased relative precision
when applying excitation equilibrium to very metal-poor stars, so that
the further investigation of non-LTE effects will be highly desirable
(Hosford et al.~in prep.). 

\subsection{The most metal-poor stars in the Galaxy}

Despite theoretical uncertainties on the exact mass range under which
the first
stars formed, it is likely that the most metal-poor objects currently
observed in the Milky Way halo are second generation stars. In case of
dwarfs/subgiants, their
abundance patterns carry direct information on the first stars
ever formed in the Galaxy \citep[e.g.][]{frebel05:nature} and/or on
still poorly known long time-scale processes which might take place
below the surface or deep into stellar interior \citep[e.g.][]{venn08,korn09}.

Determining their effective temperature and evolutionary status
(i.e.~$\logg$) is crucial to derive reliable abundances and constrain 
different scenarios. At the same time, such a quest is in stark
contrast with the many practical limitations associated with 
hyper-metal-poor stars: parallaxes are not available to help
constrain their surface gravities and even when spectra with sufficient
resolution and S/N are obtained, the model atmospheres used for the
analysis are not yet fully tested at such low metallicities. Rigorous
analyses should also take into account 3D \citep{frebel08} and NLTE 
\citep{aoki06} effects, which are expected to be considerable in this regime. 
Determining $\teff$ in a way mostly unaffected by the above
limitations is not only desirable, but also necessary to put
spectroscopic analyses on firmer grounds.  

\subsubsection{HE1327-2326}\label{he1327}

For this star the IRFM returns $\teff=6250 \pm 60$\,K in agreement 
within the errors with the spectroscopic value of $6120\pm150$\,K
obtained from the NLTE analysis of the Balmer lines \citep{korn09},
roughly with an offset of the same order of that discussed in Section 
\ref{Balmer}. 
As we already pointed out, the IRFM depends only
weakly on the adopted surface gravity: changing it by $\pm0.5$~dex
affects $\teff$ by approximately $\pm 25$\,K. In our case, we used
$\logg=3.7$ as recently determined by \citet{korn09}. The exact
metallicity of HE1327-2326 is also uncertain: although it is well
established that its $\feh < -5.0$, estimates range from $-5.9$
to $-5.4$ depending on the adopted stellar parameters and 1D/3D
LTE/NLTE analysis performed \citep{aoki06,frebel08}.
The IRFM is known to depend very little on the metallicity and we
verified this being particularly true (at least in this $\teff$
regime) for the featureless spectra of this hyper-metal-poor star:
increasing $\feh$ by $1$~dex in the IRFM affects the derived $\teff$
by less then $10$\,K. This conclusion supports the suggestion that for 
hyper-metal-poor stars colour--temperature calibration of normal
very-metal-poor stars can be used instead (see discussion in Section 
\ref{colte}).

When running the IRFM for this star we used the new grid of MARCS
model atmosphere \citep{gustafsson08} which extend down to
$\feh=-5.0$ and this value was used in our implementation. Because of the weak
metallicity dependence discussed above, very similar results are
obtained if the \citet{castelli04} grid (which stops to
$\rm{[M/H]}=-4.0$) is used instead. For the sake of ensuring our
results do not depend too much on the adopted spectra library, we also 
checked that for stars with higher metallicities MARCS or ATLAS9 
models return very similar results, with differences usually well
within $10$\,K and at most of order $20$\,K \citep[see also][]{casagrande06}. 

We feel the major source of possible systematic error stems from
reddening, which is very high for this star. We used $E(B-V)=0.076$
based on both extinction maps and interstellar absorption lines
\citep{aoki06,beers07} but it should be kept
in mind that a change of $\pm0.01$mag.~in $E(B-V)$ affects $\teff$ by 
$\pm50$\,K. 
\begin{figure}
\includegraphics[width=9.0cm]{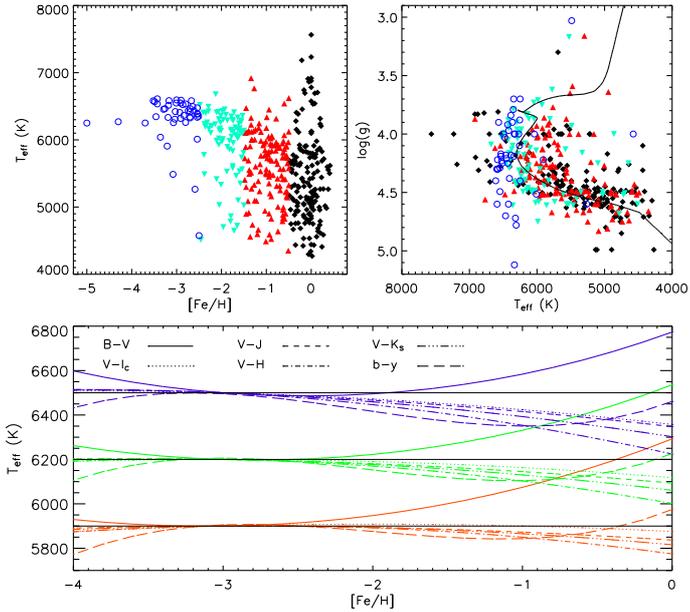}
\caption{Upper left panel: metallicities and effective temperatures of our 
sample. All stars have 2MASS and Johnson-Cousins photometry.
Upper right panel: effective temperatures and gravities of our sample. 
Symbols for different metallicity bins are the same as in the left panel. 
Overplotted for reference is a 3~Gyr solar isochrone from \citet{bertelli08}.
Lower panel: metallicity sensitivity of our colour-temperature calibration in 
different bands for stars having $\teff=6500$~K (top), $6200$~K (middle) 
and $5900$~K (lower) at $\feh=-3.0$.}
\label{f:space}
\end{figure}

\begin{figure*}
\includegraphics[scale=0.9]{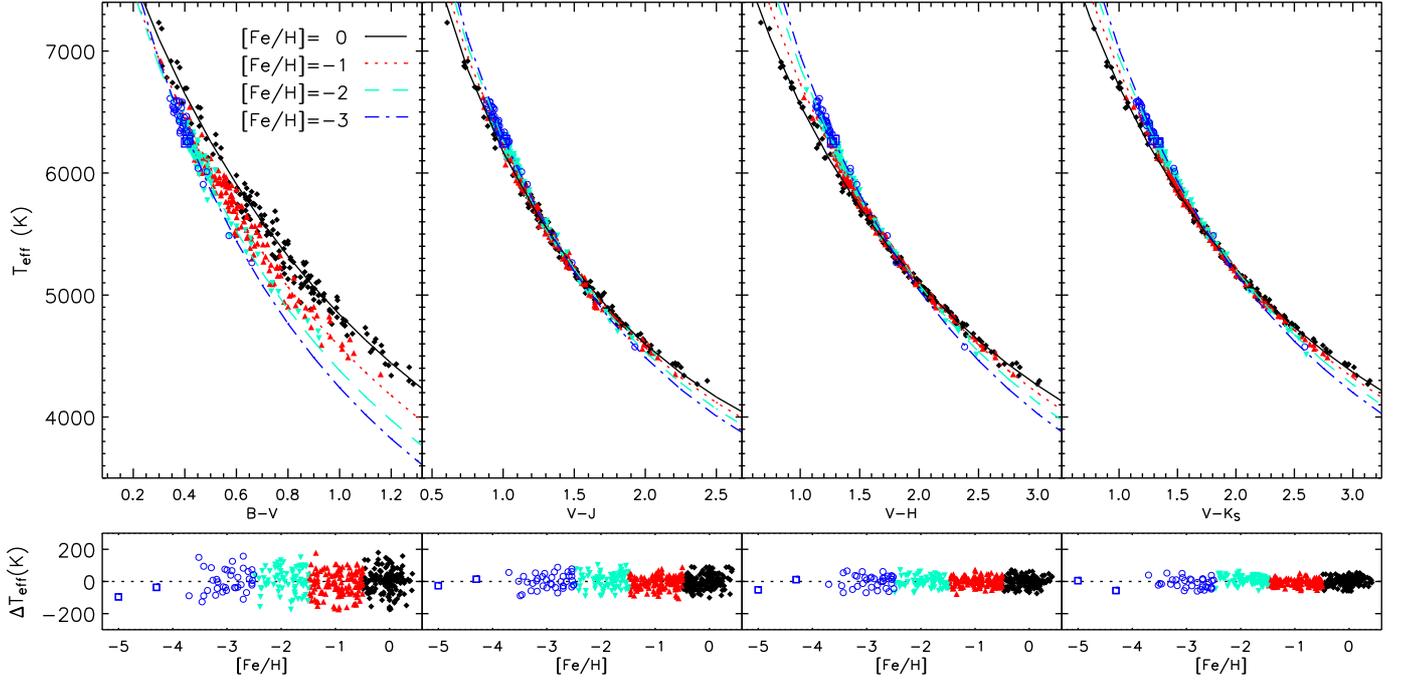}
\caption{Upper panels: empirical colour--temperature--metallicity calibrations in the metallicity bins $-0.5 < \feh \le 0.5$ (filled diamonds), $-1.5 < \feh \le -0.5$ (upward triangles), $-2.5 < \feh \le -1.5$ (downward triangles) and $\feh \le -2.5$ (open circles). Open squares are for the hyper metal-poor stars HE0233-0343 and HE1327-2326. Lower panels: residual of the fit as function of metallicity. For the two hyper-metal-poor stars, the residual is with respect to the fit at $\feh = -3.5$.}
\label{f:colte}
\end{figure*}

\subsubsection{HE0233-0343}

Though the exact metallicity of this star is still uncertain, it seems
well secured as having $\feh \lesssim -4.0$ \citep[][Garc\'ia P\'erez
  private communication]{ana08}. Its evolutionary status is also
ambiguous, with spectroscopic estimates of $\logg$ varying from
$3.5$ to $4.5$. Also in this case, the exact values of $\logg$ and
$\feh$ are not crucial for the IRFM and we checked that changing them
even considerably affects $\teff$ by an amount similar to that
discussed for HE1327-2326. We adopt $\feh=-4.0$ and $\logg=4.0$ from
which we derive $\teff=6270\pm80$\,K, without accounting for possible
systematics arising from $E(B-V)=0.025$ \citep{beers07}. As we point out in
Section \ref{colte} there might be some issue with the $R_C$ photometry
for this star. Were we to exclude this band when running the
IRFM, $\teff$ would increase by $25-35$\,K depending on the surface 
gravity assumed. Spectroscopic $\teff$ estimates for this star are
still uncertain, primarily because of its uncertain $\logg$. Were its
subgiant status to be confirmed, our effective temperature
would be in good agreement with the spectroscopic one \citep{ana08}.

\section{Empirical calibrations}\label{colte}

The effective temperatures and the bolometric luminosities derived via
IRFM for our sample allow us to build calibrations relating those
quantities to the measured colours and metallicities. As discussed in
Section \ref{compare}, to correctly account for reddening is crucial
though fortunately, for the sake of deriving colour relations,
reddening affects both the observed photometry and the derived
fundamental stellar parameters, thus making such relations --built
using dereddened colours-- independent on the adopted $E(B-V)$ in
first approximation.  

In the following we give the functional form of these calibrations,
together with the number of stars used, the standard deviation
obtained in the fitting process and the range of applicability. The results 
presented here usually match \citet{casagrande06} within the limits of those 
calibrations, but extend over a wider range now and thus supersede the previous 
work. 
Though our sample has been assembled explicitly to cover a parameter space as 
large as
possible in effective temperature and metallicity, the detection and
observation of stars with $\feh \lesssim -2.5$ is still strongly
biased around $\teff \sim 6500$\,K. Even if the formal range of
applicability of the calibrations extend well below $\feh <-3$, the
number of known metal-poor stars considerably decreases as one moves away
from the aforementioned $\teff$ (see Figure \ref{f:space}). In
particular, for metallicities below $-4$, only two stars are currently
known, a number clearly inadequate to give fits. Fortunately, at these 
temperatures calibrations at about $-3.5$ seem adequate for
even more metal-poor stars, as we discuss further in Section
\ref{tempcal} and \ref{fluxcal}. Nonetheless, we advocate particular
caution when using these calibrations in poorly sampled regions of
Figure \ref{f:space}. On the contrary for $\feh \gtrsim -2$, typical
for most of the stellar population observed in the solar neighbourhood
and Galactic star clusters, our calibrations are robust and can be
readily used for a number of purposes. 

The core of the present work is to accurately define the zero point of the 
temperature scale in many standard photometric systems; we caution however 
that in some cases real systems might not exactly reproduce standard systems, 
especially in the case of the faintest sources \citep{bessell05}. 
Users of our calibrations should always keep this 
in mind: although the zero point of the $\teff$ scale is now well defined, 
in gathering photometry from heterogeneous sources there might be small zero 
point issues between different authors, and this observational uncertainty 
--if present-- will introduce small systematic errors to our accurate empirical 
calibrations.

\subsection{Colour--Temperature--Metallicity}\label{tempcal}

To reproduce the observed $\teff$ versus colour relation and take into account the effects of metallicity, the usual fitting formula has been adopted \citep[e.g.][]{alonso96:teff_scale,ramirez05b,casagrande06,gonzalez09}
\begin{equation}\label{eq:ct}
\theta_{\rm{eff}}=a_0 + a_1 X + a_2 X^2 + a_3 X \feh + a_4 \feh + a_5 \feh^2
\end{equation}
where $\theta_{\rm{eff}}=5040/ \teff$, $X$ represents the colour and 
$a_i\,(i=0,\ldots,5)$ are the coefficients of the fit obtained
iteratively, discarding points departing more than $3 \sigma$.  
\begin{table*}
\centering
\caption{Coefficients and range of applicability of the
  colour--temperature--metallicity relations. The photometric systems
  are Johnson-Cousins $BV(RI)_C$, 2MASS $JHK_S$, Tycho2 $(BV)_T$ and 
  Str\"omgren $by$. For the latter, additional corrections as function of 
  $\feh$ and $(b-y)$ apply, as discussed in Section \ref{stromgren}.
  For some indices the calibrations are given down to
  $\feh=-5.0$, meaning that the effective temperatures of such a
  metal-poor star can be recovered using $\feh=-3.5$ in
  Eq.~(\ref{eq:ct}). Notice that only two hyper metal-poor stars are
  currently known and caution should be used, as discussed in the text. 
  Especially for metal-poor stars, please refer to Figure 
  \ref{f:space} to check that the calibration is not extrapolated 
  outside its $\feh$ range.}
\label{tab:ct}
\begin{tabular}{lcccccccccc}
\hline\hline
Colour    & $\feh$ range & Colour range  & $a_0$    & $a_1$    & $a_2$     & $a_3$     & $a_4$ & $a_5$ & $N$ & $\sigma (\teff)$ \\     
\hline
$B-V$     & $[-5.0,0.4]$ & $[0.18,1.29]$ & $0.5665$ & $0.4809$ & $-0.0060$ & $-0.0613$  & $-0.0042$ & $-0.0055$ & 400 & 73\\     
$V-R_C$   & $[-5.0,0.3]$ & $[0.24,0.80]$ & $0.4386$ & $1.4614$ & $-0.7014$ &  $-0.0807$ & $0.0142$  & $-0.0015$ & 201 & 62\\
$(R-I)_C$ & $[-5.0,0.3]$ & $[0.23,0.68]$ & $0.3296$ & $1.9716$ & $-1.0225$ & $-0.0298$  & $0.0329$  & $0.0035$  & 211 & 82\\
$V-I_C$   & $[-5.0,0.3]$ & $[0.46,1.47]$ & $0.4033$ & $0.8171$ & $-0.1987$ & $-0.0409$  & $0.0319$  & $0.0012$  & 208 & 59\\
$V-J$     & $[-5.0,0.4]$ & $[0.61,2.44]$ & $0.4669$ & $0.3849$ & $-0.0350$ & $-0.0140$  & $0.0225$  & $0.0011$  & 401 & 42\\
$V-H$     & $[-5.0,0.4]$ & $[0.67,3.01]$ & $0.5251$ & $0.2553$ & $-0.0119$ & $-0.0187$  & $0.0410$  & $0.0025$  & 401 & 33\\
$V-K_S$   & $[-5.0,0.4]$ & $[0.78,3.15]$ & $0.5057$ & $0.2600$ & $-0.0146$ &  $-0.0131$ & $0.0288$  & $0.0016$  & 401 & 25\\
$J-K_S$   & $[-5.0,0.4]$ & $[0.07,0.80]$ & $0.6393$ & $0.6104$ & $0.0920$  & $-0.0330$  & $0.0291$  & $0.0020$  & 412 & 132\\
$(B-V)_T$ & $[-2.7,0.4]$ & $[0.19,1.49]$ & $0.5839$ & $0.4000$ & $-0.0067$ & $-0.0282$  & $-0.0346$ & $-0.0087$ & 251 & 79\\
$V_T-J$   & $[-2.7,0.4]$ & $[0.77,2.56]$ & $0.4525$ & $0.3797$ & $-0.0357$ & $-0.0082$  & $0.0123$  & $-0.0009$ & 272 & 43\\
$V_T-H$   & $[-2.7,0.4]$ & $[0.77,3.16]$ & $0.5286$ & $0.2354$ & $-0.0073$ &  $-0.0182$ & $0.0401$  & $0.0021$  & 263 & 26\\
$V_T-K_S$ & $[-2.4,0.4]$ & $[0.99,3.29]$ & $0.4892$ & $0.2634$ & $-0.0165$ & $-0.0121$  & $0.0249$  & $-0.0001$ & 258 & 18\\
$b-y$     & $[-3.7,0.5]$ & $[0.18,0.72]$ & $0.5796$ & $0.4812$ & $0.5747$  & $-0.0633$  & $0.0042$  & $-0.0055$ & 1120 & 62 \\
\hline 
\end{tabular}
\begin{list}{}{}
\item[] $N$ is the number of stars employed for the fit after the $3 \sigma$
clipping and $\sigma(\teff)$ is the standard deviation (in Kelvin) of the
proposed calibrations. Notice that the standard deviation does not account for 
the uncertainty in the zero point of the temperature scale, which is of order 
$15-20$\,K (Section \ref{tuning} and \ref{inter}).
\end{list}
\end{table*}

The IRFM depends only very mildly on the adopted $\logg$ (Section 
\ref{proscons}) but certain colours could be more affected: for all 
indices we have checked the residual of our calibration and did not 
find any obvious trend with $\logg$. Nevertheless, a dependence 
on the gravity could be built into the calibrations, since $\logg$ 
decreases as one moves from cool dwarfs to hotter turn-off stars 
(Figure \ref{f:space}). 

The coefficients for various colour indices are given with their range of 
applicability in Table \ref{tab:ct} and a comparison between the polynomial 
fits and our sample of stars is shown in Figure \ref{f:colte}. 
We remark that the functional form of Eq. (\ref{eq:ct}) may 
return non-physical values when extrapolated to very low 
metallicities, as extensively discussed by \citet{ryan99} for the 
calibration of \citet{alonso96:teff_scale} below $\feh \sim -2.5$. 
We have considerably increased the number of very metal-poor 
(turnoff) stars and our calibration behaves as one would 
expect, i.e.\ it shows a decreasing sensitivity on $\feh$ when moving 
from $-2$ to $-3$, where the metallicity sensitivity vanishes in 
all bands (Figure \ref{f:space}). Moving to $\feh=-4$ (or lower), the 
diverging behaviour in Figure \ref{f:space} reflects the form of the 
fitting function and the values of the coefficients rather than 
the characteristics of metal-poor turnoff stars.
In Figure \ref{f:colte} the two hyper metal-poor stars (represented by open squares) clearly follow the same trend of other iron deficient stars with similar effective temperatures. Using Eq.~(\ref{eq:ct}) at a fixed $\feh=-3.5$ recovers their IRFM $\teff$ within the typical accuracy of the calibration. This is always true for HE1327-2326, and also for HE0233-0343 except when using the $R_C$ index, possibly indicating a photometric issue in this band for the latter star. This comparison thus warrants the applicability of our calibrations for hyper-metal-poor stars if $\feh=-3.5$ is assumed and a typical $\teff \sim 6200$\,K is obtained. How well this holds at other effective temperatures is still unknown. 

The calibration presented here applies till late K-type dwarfs. Those 
interested in M dwarfs, can instead refer to \citet{casagrande08}: though in 
that work the zero point has not been constrained using solar twins, the 
absolute calibration adopted was similar to that used here, resulting in 
effective temperatures approximately on the same scale. Nonetheless, if a 
link between the two scales is needed, we advise users to a careful 
case-by-case study, also considering that the calibration for M dwarfs has a 
different functional form and does not include any metallicity term.

\subsubsection{Str\"omgren calibration}\label{stromgren}

The Str\"omgren index $b-y$ deserves a separate discussion. It is 
often used as a $\teff$ indicator, but because of its very nature has 
a strong sensitivity on the metallicity and a proper functional form 
is not trivial. \citet{alonso96:teff_scale} excluded the coolest 
dwarfs, where the dependence of $b-y$ upon $\teff$ 
possibly flattens out. Yet, for the most metal poor stars that 
calibration diverges to unphysical values, as discussed in 
\citet{ryan99}.

For $b-y$ we have verified that a calibration of the form of 
Eq. (\ref{eq:ct}) has strong residuals as function of both colour 
and metallicity and used polynomial fits to correct such trends, 
i.e.\ $\teff = 5040/ \theta_{\rm{eff}} + P(\feh,b-y)$.
To this purpose, we have increased the sample with more than 
1000 stars from the GCS catalogue \citep{nordstrom04} all having 
Str\"omgren photometry, spectroscopic metallicities from an updated 
version of the Cayrel catalogue (Mel\'endez, in prep.) and for 
which the IRFM could be applied directly using Tycho2 and 2MASS 
(Casagrande et al.~in prep.). 

\begin{figure}
\includegraphics[width=9.0cm]{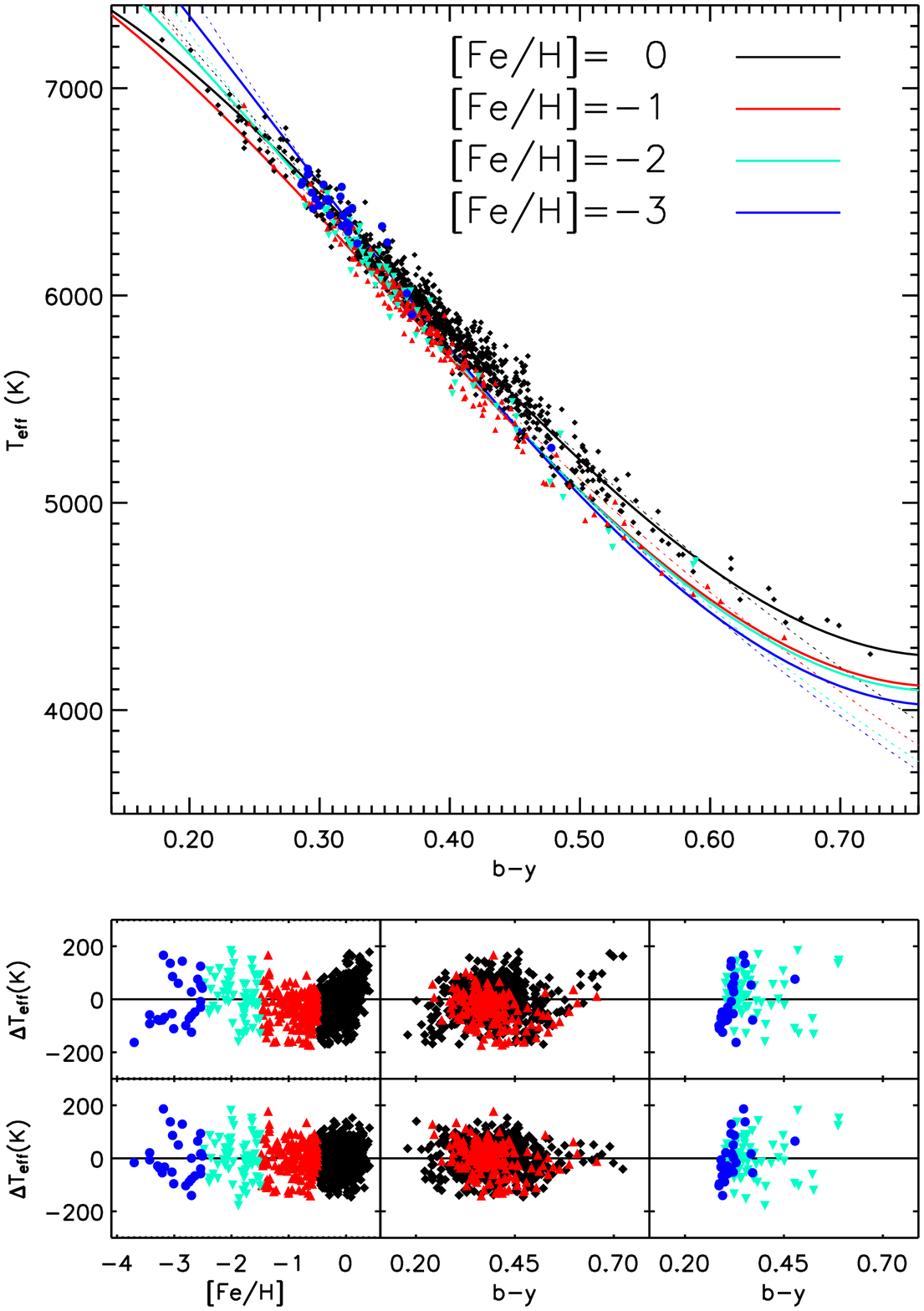}
\caption{Upper panel: empirical colour-temperature-metallicity calibration in 
$b-y$ before (dotted) and after (continuous lines) the polynomial correction.
Central and lower panels: residuals before and after the polynomial 
corrections.}
\label{f:stromgren}
\end{figure}

We checked that a third order polynomial in both colour and 
metallicity was enough; the calibration before and after adopting 
such a correction is shown in Figure \ref{f:stromgren} and the 
coefficients, given in the form 
$P(\feh,b-y)=\sum_{i=0}^{3}M_{i}\feh^{i}+\sum_{i=0}^{3}C_{i} (b-y)^{i}$ 
are
$M_{0}=-1.9$, $M_{1}=130.4$, $M_{2}=125.7$, $M_{3}=27.4$, 
$C_{0}=-1003.7$, $C_{1}=7325.9$, $C_{2}=-17207.4$, $C_{3}=12977.7$.
Notice that the form of these corrections can lead to unphysical 
values if extrapolated and should never be applied outside of the 
colour and [Fe/H] ranges of Figure \ref{f:stromgren}.

\begin{figure*}
\includegraphics[scale=0.9]{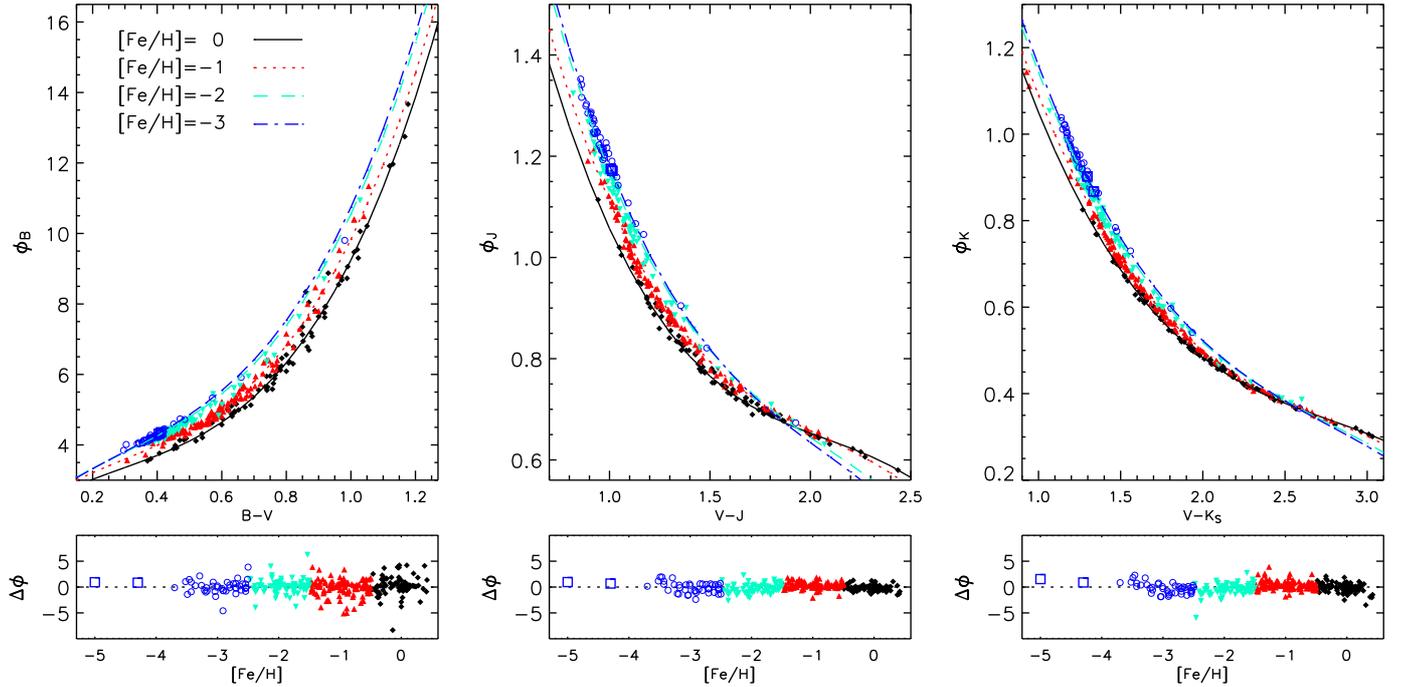}
\caption{Same as Figure \ref{f:colte}, but for the
  colour--flux--metallicity calibrations. The reduced flux in
  different bands $\phi_{\zeta} =
  \mathcal{F_{\rm{Bol}}}(\rm{Earth})\,10^{0.4\,m_{\zeta}}$ is plotted
  as function of different colour indices in units of 
  $10^{-0.5} \rm{erg\,cm^{-2}\,s^{-1}}$.}
\label{f:colflu}
\end{figure*}

\subsection{Colour--Flux--Metallicity}\label{fluxcal}

We adopt the same definition of \citet{casagrande06} to define the bolometric correction in a given $\zeta$ band, where
\begin{equation}\label{eq:bc}
BC_{\zeta}=m_{\rm{Bol}}-m_{\zeta}
\end{equation}
and the zero point of the $m_{\rm{bol}}$ scale is fixed by choosing $M_{\rm{Bol},\odot}=4.74$. Empirical bolometric corrections in various bands can thus be readily computed using Eq.~(\ref{eq:bc}) and dereddening the observed magnitudes given in Table 8.
\begin{table*}
\centering
\caption{Coefficients and range of applicability of the flux
  calibrations for various $\phi_{\zeta} =
  \mathcal{F_{\rm{Bol}}}(\rm{Earth})\,10^{0.4\,m_{\zeta}}$.}
\label{tab:cf}
\begin{tabular}{llccccccccccc}
\hline\hline
$\phi_{\zeta}$ & Colour & $\feh$ range & Colour range & $b_0$ & $b_1$ & $b_2$ & $b_3$ & $b_4$ & $b_5$ & $b_6$ & $N$ & $\sigma(\%)$ \\     
\hline
$B_T$ & $(B-V)_T$ & $[-2.7,0.4]$ &  $[0.19,1.43]$ &   $2.1904$ &  $5.7106$   & $-6.7110$   & $7.4160$    & $-0.6704$   & $-0.1501$   & $-0.0720$   &  $260$  & $3.1$\\
$B_T$ & $V_T-J$   & $[-2.7,0.4]$ &  $[0.51,2.56]$ &   $1.8160$ &  $3.2833$   & $-2.3210$   & $1.7358$    & $1.2140$    & $-1.0830$   & $0.0343$    &  $261$  & $4.2$\\
$B_T$ & $V_T-H$   & $[-2.7,0.4]$ &  $[0.53,3.16]$ &   $1.7597$ &  $3.1896$   & $-1.8419$   & $0.9465$    & $0.9826$    & $-0.9055$   & $0.0809$    &  $255$  & $4.0$\\
$B_T$ & $V_T-K_S$ & $[-2.7,0.4]$ &  $[0.59,3.29]$ &   $1.7202$ &  $3.0146$   & $-1.6377$   & $0.8033$    & $0.8591$    & $-0.8644$   & $0.0669$    &  $262$  & $3.9$\\
$V_T$ & $(B-V)_T$ & $[-2.7,0.4]$ &  $[0.19,1.43]$ &   $2.7098$ &  $-0.2765$  & $0.1523$    & $0.8122$    & $-0.2261$   & $-0.1789$   &  $-0.0413$  & $253$   & $2.7$\\
$V_T$ & $V_T-J$    & $[-2.7,0.4]$ &  $[0.62,2.53]$ &   $2.1815$ &  $0.9268$   & $-0.7701$   & $0.4029$    & $0.1047$    & $-0.2609$   &  $-0.0048$  & $249$   & $0.7$\\
$V_T$ & $V_T-H$    & $[-2.7,0.4]$ &  $[0.68,3.16]$ &   $2.1800$ &  $0.8514$   & $-0.5793$   & $0.2235$    & $0.0936$    & $-0.2458$   &  $0.0019$   & $261$   & $0.9$\\
$V_T$ & $V_T-K_S$  & $[-2.7,0.4]$ &  $[0.59,3.29]$ &   $2.2565$ &  $0.6787$   & $-0.4536$   & $0.1800$    & $0.0785$    & $-0.2407$   &  $-0.0011$  & $256$   & $0.9$\\ 
$B$   & $B-V$     & $[-5.0,0.4]$ &  $[0.18,1.22]$ &   $1.9571$ &  $  6.9680$ & $ -11.0277$ & $  11.4450$ &   $-0.4975$ &   $-0.1276$ &   $-0.0432$ &   $331$ & $2.3$\\
$B$   & $V-R_C$   & $[-5.0,0.3]$ &  $[0.24,0.79]$ &   $2.0002$ &  $  6.6483$ & $  -4.6407$ & $  25.3881$ &   $ 0.9547$ &   $-0.3756$ &   $-0.0067$ &   $186$ & $1.9$\\
$B$   & $(R-I)_C$ & $[-5.0,0.3]$ &  $[0.23,0.68]$ &   $9.8257$ &  $-57.0297$ & $ 152.2749$ & $ -77.6378$ &   $ 4.3253$ &   $-1.1377$ &   $ 0.0411$ &   $202$ & $3.2$\\
$B$   & $V-I_C$   & $[-5.0,0.3]$ &  $[0.46,1.47]$ &   $4.3948$ &  $ -6.0713$ & $   9.6862$ & $   0.2327$ &   $ 0.9298$ &   $-0.5392$ &   $ 0.0089$ &   $196$ & $1.6$\\
$B$   & $V-J$     & $[-5.0,0.4]$ &  $[0.50,2.44]$ &   $1.6664$ &  $  3.5465$ & $  -2.5257$ & $   1.5310$ &   $ 0.4259$ &   $-0.4354$ &   $ 0.0047$ &   $332$ & $2.8$\\
$B$   & $V-H$     & $[-5.0,0.4]$ &  $[0.52,2.84]$ &   $1.6852$ &  $  3.2925$ & $  -1.9206$ & $   0.8026$ &   $ 0.2172$ &   $-0.1301$ &   $ 0.0346$ &   $328$ & $2.9$\\
$B$   & $V-K_S$   & $[-5.0,0.4]$ &  $[0.57,3.03]$ &   $1.5185$ &  $  3.3566$ & $  -1.8830$ & $   0.7301$ &   $ 0.2887$ &   $-0.2929$ &   $ 0.0240$ &   $363$ & $2.8$\\
$V$   & $B-V$     & $[-5.0,0.4]$ &  $[0.30,1.03]$ &   $1.2581$ &  $  5.8828$ & $  -9.9287$ & $   6.8432$ &   $ 0.2290$ &   $-0.3935$ &   $-0.0420$ &   $241$ & $1.9$\\
$V$   & $V-R_C$   & $[-5.0,0.3]$ &  $[0.24,0.79]$ &   $2.6659$ &  $ -1.6396$ & $   3.9243$ & $   2.9911$ &   $ 0.0978$ &   $-0.2339$ &   $-0.0252$ &   $177$ & $0.7$\\
$V$   & $(R-I)_C$ & $[-5.0,0.3]$ &  $[0.25,0.68]$ &   $4.9994$ &  $-20.1727$ & $  49.0418$ & $ -27.5918$ &   $ 0.9465$ &   $-0.4491$ &   $-0.0166$ &   $197$ & $1.5$\\
$V$   & $V-I_C$   & $[-5.0,0.3]$ &  $[0.48,1.47]$ &   $3.4468$ &  $ -3.8760$ & $   4.5692$ & $  -0.7285$ &   $ 0.1832$ &   $-0.2991$ &   $-0.0231$ &   $184$ & $0.8$\\
$V$   & $V-J$     & $[-5.0,0.4]$ &  $[0.73,2.21]$ &   $1.8195$ &  $  1.5562$ & $  -1.3322$ & $   0.5627$ &   $ 0.1249$ &   $-0.3112$ &   $-0.0213$ &   $314$ & $0.8$\\
$V$   & $V-H$     & $[-5.0,0.4]$ &  $[0.67,3.01]$ &   $2.0139$ &  $  1.0845$ & $  -0.8071$ & $   0.2761$ &   $ 0.0567$ &   $-0.2147$ &   $-0.0124$ &   $369$ & $0.9$\\
$V$   & $V-K_S$   & $[-5.0,0.4]$ &  $[0.93,3.15]$ &   $1.7662$ &  $  1.4154$ & $  -0.9302$ & $   0.2726$ &   $ 0.0692$ &   $-0.2506$ &   $-0.0160$ &   $316$ & $0.9$\\
$R_C$ & $B-V$     & $[-5.0,0.3]$ &  $[0.35,1.29]$ &   $2.5759$ &  $ -1.8536$ & $   1.3042$ & $   0.1015$ &   $-0.0130$ &   $-0.1229$ &   $-0.0142$ &   $179$ & $0.8$\\
$R_C$ & $V-R_C$   & $[-5.0,0.3]$ &  $[0.24,0.79]$ &   $2.7031$ &  $ -4.2859$ & $   6.9274$ & $  -2.1959$ &   $ 0.1482$ &   $-0.1968$ &   $-0.0186$ &   $180$ & $0.7$\\
$R_C$ & $(R-I)_C$ & $[-5.0,0.3]$ &  $[0.23,0.68]$ &   $3.2131$ &  $ -8.5410$ & $  17.3691$ & $  -9.1350$ &   $ 0.4602$ &   $-0.3054$ &   $-0.0171$ &   $203$ & $1.0$\\
$R_C$ & $V-I_C$ & $[-5.0,0.3]$ &  $[0.48,1.47]$ &   $2.9759$ &  $ -3.2013$ & $   2.9454$ & $  -0.6516$ &   $ 0.1331$ &   $-0.2408$ &   $-0.0187$ &   $185$ & $0.7$\\
$R_C$ & $V-J$ & $[-5.0,0.3]$ &  $[0.86,2.36]$ &   $2.5806$ &  $ -1.0234$ & $   0.4055$ & $   0.0107$ &   $ 0.0874$ &   $-0.2508$ &   $-0.0181$ &   $184$ & $0.7$\\
$R_C$ & $V-H$ & $[-5.0,0.3]$ &  $[0.93,2.99]$ &   $2.5007$ &  $ -0.6801$ & $   0.1842$ & $   0.0176$ &   $ 0.0746$ &   $-0.2604$ &   $-0.0185$ &   $196$ & $0.7$\\
$R_C$ & $V-K_S$ & $[-5.0,0.3]$ &  $[1.00,3.13]$ &   $2.5606$ &  $ -0.7448$ & $   0.2212$ & $   0.0049$ &   $ 0.0665$ &   $-0.2509$ &   $-0.0184$ &   $193$ & $0.7$\\
$I_C$ & $B-V$ & $[-5.0,0.3]$ &  $[0.35,1.29]$ &   $2.6765$ &  $ -3.8643$ & $   3.7834$ & $  -1.2273$ &   $ 0.0145$ &   $-0.0358$ &   $-0.0015$ &   $200$ & $1.0$\\
$I_C$ & $V-R_C$ & $[-5.0,0.3]$ &  $[0.24,0.79]$ &   $2.6963$ &  $ -6.8081$ & $  11.3579$ & $  -6.1859$ &   $ 0.1798$ &   $-0.1418$ &   $-0.0112$ &   $191$ & $1.1$\\
$I_C$ & $(R-I)_C$ & $[-5.0,0.3]$ &  $[0.27,0.68]$ &   $3.1500$ &  $-10.0132$ & $  18.2682$ & $ -10.8668$ &   $ 0.3653$ &   $-0.2329$ &   $-0.0123$ &   $177$ & $0.9$\\
$I_C$ & $V-I_C$ & $[-5.0,0.3]$ &  $[0.48,1.47]$ &   $3.0203$ &  $ -4.5225$ & $   4.0375$ & $  -1.1781$ &   $ 0.1371$ &   $-0.1866$ &   $-0.0122$ &   $197$ & $0.8$\\
$I_C$ & $V-J$ & $[-5.0,0.3]$ &  $[0.86,2.36]$ &   $2.7912$ &  $ -2.2548$ & $   1.1687$ & $  -0.1986$ &   $ 0.0817$ &   $-0.1908$ &   $-0.0118$ &   $186$ & $1.1$\\
$I_C$ & $V-H$ & $[-5.0,0.3]$ &  $[0.93,2.99]$ &   $2.7888$ &  $ -1.8271$ & $   0.7688$ & $  -0.1061$ &   $ 0.0734$ &   $-0.2132$ &   $-0.0138$ &   $187$ & $1.0$\\
$I_C$ & $V-K_S$ & $[-5.0,0.3]$ &  $[1.00,3.13]$ &   $2.7797$ &  $ -1.7014$ & $   0.6710$ & $  -0.0868$ &   $ 0.0603$ &   $-0.1891$ &   $-0.0123$ &   $193$ & $0.9$\\
$J$ & $B-V$ & $[-5.0,0.4]$ &  $[0.30,1.29]$ &   $2.2253$ &  $ -3.5932$ & $   2.9303$ & $  -0.8741$ &   $ 0.0199$ &   $ 0.0132$ &   $ 0.0057$ &   $346$ & $3.4$\\
$J$ & $V-R_C$ & $[-5.0,0.3]$ &  $[0.24,0.79]$ &   $2.5765$ &  $ -8.1969$ & $  12.1713$ & $  -6.3037$ &   $ 0.1393$ &   $-0.0769$ &   $-0.0048$ &   $186$ & $2.9$\\
$J$ & $(R-I)_C$ & $[-5.0,0.3]$ &  $[0.27,0.68]$ &   $2.9723$ &  $-10.6481$ & $  16.3430$ & $  -8.5334$ &   $ 0.2971$ &   $-0.1854$ &   $-0.0095$ &   $195$ & $3.0$\\
$J$ & $V-I_C$ & $[-5.0,0.3]$ &  $[0.52,1.47]$ &   $2.8966$ &  $ -5.1154$ & $   4.0119$ & $  -1.0879$ &   $ 0.1059$ &   $-0.1232$ &   $-0.0066$ &   $192$ & $2.6$\\
$J$ & $V-J$ & $[-5.0,0.4]$ &  $[0.82,2.44]$ &   $2.7915$ &  $ -2.8096$ & $   1.2799$ & $  -0.2049$ &   $ 0.0479$ &   $-0.1059$ &   $-0.0054$ &   $308$ & $0.8$\\
$J$ & $V-H$ & $[-5.0,0.4]$ &  $[0.88,2.99]$ &   $2.5885$ &  $ -2.0262$ & $   0.7430$ & $  -0.0963$ &   $ 0.0587$ &   $-0.1577$ &   $-0.0092$ &   $303$ & $1.9$\\
$J$ & $V-K_S$ & $[-5.0,0.4]$ &  $[0.93,3.13]$ &   $2.5578$ &  $ -1.8710$ & $   0.6433$ & $  -0.0785$ &   $ 0.0457$ &   $-0.1326$ &   $-0.0078$ &   $314$ & $1.7$\\
$H$ & $B-V$ & $[-5.0,0.4]$ &  $[0.18,1.29]$ &   $2.1337$ &  $ -3.6473$ & $   2.6261$ & $  -0.6782$ &   $-0.0780$ &   $ 0.1274$ &   $ 0.0179$ &   $331$ & $4.6$\\
$H$ & $V-R_C$ & $[-5.0,0.3]$ &  $[0.24,0.79]$ &   $2.5341$ &  $ -8.8850$ & $  12.8801$ & $  -6.5281$ &   $ 0.0339$ &   $-0.0012$ &   $ 0.0022$ &   $184$ & $3.6$\\
$H$ & $(R-I)_C$ & $[-5.0,0.3]$ &  $[0.23,0.68]$ &   $2.9097$ &  $-11.1909$ & $  16.5901$ & $  -8.3344$ &   $ 0.1844$ &   $-0.1169$ &   $-0.0047$ &   $192$ & $3.6$\\
$H$ & $V-I_C$ & $[-5.0,0.3]$ &  $[0.48,1.47]$ &   $2.8833$ &  $ -5.5447$ & $   4.2495$ & $  -1.1267$ &   $ 0.0504$ &   $-0.0512$ &   $-0.0009$ &   $195$ & $3.1$\\
$H$ & $V-J$ & $[-5.0,0.4]$ &  $[0.50,2.44]$ &   $2.5764$ &  $ -2.6119$ & $   1.0580$ & $  -0.1490$ &   $ 0.0033$ &   $-0.0098$ &   $ 0.0031$ &   $353$ & $2.8$\\
$H$ & $V-H$ & $[-5.0,0.4]$ &  $[0.88,3.01]$ &   $2.4574$ &  $ -2.0093$ & $   0.6768$ & $  -0.0808$ &   $ 0.0246$ &   $-0.0665$ &   $-0.0016$ &   $344$ & $1.0$\\
$H$ & $V-K_S$ & $[-5.0,0.4]$ &  $[0.57,3.15]$ &   $2.3732$ &  $ -1.7778$ & $   0.5485$ & $  -0.0599$ &   $ 0.0140$ &   $-0.0407$ &   $ 0.0005$ &   $363$ & $2.2$\\
$K_S$ & $B-V$ & $[-5.0,0.4]$ &  $[0.30,1.29]$ &   $2.1537$ &  $ -3.9640$ & $   3.0680$ & $  -0.8653$ &   $-0.0586$ &   $ 0.1098$ &   $ 0.0163$ &   $353$ & $4.9$\\
$K_S$ & $V-R_C$ & $[-5.0,0.3]$ &  $[0.24,0.76]$ &   $2.5709$ &  $ -9.5441$ & $  14.4103$ & $  -7.6430$ &   $ 0.0585$ &   $-0.0186$ &   $ 0.0006$ &   $190$ & $3.8$\\
$K_S$ & $(R-I)_C$ & $[-5.0,0.3]$ &  $[0.25,0.68]$ &   $2.8803$ &  $-11.4591$ & $  17.4060$ & $  -9.0419$ &   $ 0.2418$ &   $-0.1469$ &   $-0.0066$ &   $199$ & $3.6$\\
$K_S$ & $V-I_C$ & $[-5.0,0.3]$ &  $[0.48,1.47]$ &   $2.7928$ &  $ -5.4377$ & $   4.1682$ & $  -1.1105$ &   $ 0.0661$ &   $-0.0690$ &   $-0.0019$ &   $201$ & $3.2$\\
$K_S$ & $V-J$ & $[-5.0,0.4]$ &  $[0.50,2.36]$ &   $2.6548$ &  $ -2.8832$ & $   1.2411$ & $  -0.1878$ &   $ 0.0146$ &   $-0.0315$ &   $ 0.0013$ &   $328$ & $2.9$\\
$K_S$ & $V-H$ & $[-5.0,0.4]$ &  $[0.88,2.99]$ &   $2.4939$ &  $ -2.1600$ & $   0.7593$ & $  -0.0946$ &   $ 0.0342$ &   $-0.0879$ &   $-0.0029$ &   $317$ & $2.5$\\
$K_S$ & $V-K_S$ & $[-5.0,0.4]$ &  $[0.93,3.03]$ &   $2.5097$ &  $ -2.0732$ & $   0.6972$ & $  -0.0836$ &   $ 0.0229$ &   $-0.0641$ &   $-0.0017$ &   $328$ & $1.1$\\
\hline 
\end{tabular}
\begin{list}{}{}
\item[] $N$ is the number of stars employed for the fit after the $3$ sigma
clipping and $\sigma(\%)$ is the standard deviation of the final calibrations in percent. The coefficients of the calibrations $b_i$ are given in units of $10^{-0.5} \rm{erg\,cm^{-2}\,s^{-1}}$.
\end{list}
\end{table*}

A complementary way of deriving stellar integrated flux via
photometric indices is given in the form of \citet{casagrande06}
\begin{displaymath}
\mathcal{F_{\rm{Bol}}}\rm{(Earth)} = 10^{-0.4\,m_{\zeta}} \Big( b_0 + b_1 X + b_2 X^2 + b_3 X^3 
\end{displaymath}
\begin{equation}\label{eq:cf}
\phantom{\mathcal{F_{\rm{Bol}}}\rm{(Earth)} =} + b_4 X \feh + b_5 \feh + b_6 \feh^2 \Big) .
\end{equation}

As for the temperature calibrations, also in this case the fluxes of
the two hyper metal-poor stars can be recovered adopting $\feh=-3.5$
in Eq.~(\ref{eq:cf}), though we caution that the license of this
approach for considerably bluer or redder indices is still unknown. 

\subsection{Colour-Angular diameters}

Limb-darkened angular diameters can be readily derived from the basic 
definition involving effective temperatures and bolometric
fluxes, using the calibrations given in Sections \ref{tempcal} and
\ref{fluxcal}. Nonetheless, very tight and simple relations exist in
the $J$ band and in Table \ref{tab:diam} we give them in the form of
\citet{casagrande06} 
\begin{equation}
\theta = c_0 + c_1 \sqrt{\phi(m_J,X)}
\end{equation}
where 
\begin{equation}
\phi(m_J,X)=10^{-0.4\,m_J}X
\end{equation}
for a given colour index X. These relations show remarkably small
scatter and no metallicity dependence, thus proving ideal to build a
network of small calibrators for interferometric measurements, for 
characterizing extrasolar planet transits or microlensing events. 
\begin{table*}
\centering
\caption{Coefficients and range of applicability of the angular
  diameter calibrations.}
\label{tab:diam}
\begin{tabular}{lccccc}
\hline\hline
Colour & Colour range & $c_0$ & $c_1$ & $N$ & $\sigma(\%)$ \\
\hline 
$V-J$     & $[0.73,2.44]$ & $0.00015$ & $4.65293$ & 394 & 2.2\\
$V-H$     & $[0.88,3.01]$ & $0.00004$ & $4.15613$ & 389 & 1.8\\
$V-K_S$   & $[0.93,3.15]$ & $0.00020$ & $4.05037$ & 394 & 1.9\\
$V_T-J$   & $[0.72,2.56]$ & $0.00218$ & $4.44568$ & 270 & 2.1\\
$V_T-H$   & $[0.87,3.14]$ & $0.00227$ & $3.98945$ & 255 & 1.5\\
$V_T-K_S$ & $[0.92,3.27]$ & $0.00286$ & $3.88433$ & 268 & 1.5\\
\hline
\end{tabular}
\begin{list}{}{}
\item[] $N$ is the number of stars employed for the fit after the $3 \sigma$
clipping and $\sigma(\%)$ is the standard deviation of the calibrations.
\end{list}
\end{table*}

\subsection{The colours of the Sun}

The interpolation of Eq.~(\ref{eq:ct}) at $\teff=5777$\,K and
$\feh=0$ returns the colours of the Sun, which are given in Table
\ref{tab:sun}. For the Tycho2 and 2MASS system, those can be readily
compared with the averaged ones from the twins of Section \ref{twins}:
not unexpectedly there is good agreement, all but one within few
millimag, which usually (at maximum) correspond to few ($20$) K in
$\teff$. We have also checked that fitting our twins as function of
$\teff$ and $\feh$ returns colours almost identical to their average,
further confirming that our sample of twins is homogeneously
distributed in temperature and metallicity around the colours of the Sun
inferred from our scale. 
\begin{table*}
\centering
\caption{The colours of the Sun. For the indices obtained inverting our
  $\teff$ scale, random errors are from the dispersion of the fits in 
  Table \ref{tab:ct}. The uncertainty on the zero point of our
  $\teff$ scale is of order $15$\,K (Section \ref{tuning}), which
  usually implies systematic errors considerably smaller than the
  random ones. The only exception is for optical-infrared
  indices which are very sensitive to $\teff$ and show small intrinsic
  scatter, of the order of the aforementioned zero point uncertainty.
  Also shown for comparison are the averaged colours and standard deviation of 
  the solar twins, as well as the synthetic colours computed from the ATLAS9 
  \citep{castelli04} and MARCS \citep{gustafsson08} models.}
\label{tab:sun}
\begin{tabular}{ccccc}
\hline\hline
           &    $\teff$ scale  & MARCS & ATLAS9 & Twins \\
           & $(\rm{rand.} + \rm{syst.\;errors})$ &    &        &       \\
\hline
$B-V$      & $0.641\pm 0.024 \pm 0.004$  & 0.622 & 0.645  & \\
$V-R_C$    & $0.359\pm 0.010 \pm 0.003$  & 0.357 & 0.358  & \\
$(R-I)_C$  & $0.333\pm 0.010 \pm 0.002$  & 0.347 & 0.349  & \\
$V-I_C$    & $0.690\pm 0.016 \pm 0.004$  & 0.704 & 0.707  & \\
$V-J$      & $1.180\pm 0.021 \pm 0.007$  & 1.171 & 1.180  & \\
$V-H$      & $1.460\pm 0.023 \pm 0.010$  & 1.458 & 1.479  & \\
$V-K_S$    & $1.544\pm 0.018 \pm 0.010$  & 1.543 & 1.553  & \\
$J-K_S$    & $0.362\pm 0.029 \pm 0.003$  & 0.372 & 0.373  & \\
$(B-V)_T$  & $0.730\pm 0.031 \pm 0.006$  & 0.723 & 0.750  & $0.735 \pm 0.024^{\,a}$\\
$V_T-J$    & $1.254\pm 0.022 \pm 0.008$  & 1.240 & 1.250  & $1.254 \pm 0.041^{\,a}$\\
$V_T-H$    & $1.534\pm 0.019 \pm 0.011$  & 1.527 & 1.550  & $1.549 \pm 0.048^{\,a}$\\
$V_T-K_S$  & $1.619\pm 0.013 \pm 0.011$  & 1.612 & 1.623  & $1.623 \pm 0.040^{\,a}$\\
$b-y$      & $0.409\pm 0.010 \pm 0.002$  &       &        & $0.409 \pm 0.003^{\,b}$\\
\hline 
\end{tabular}
\begin{list}{}{}
\item[] $^{a}$ Average and standard deviation of the colours in 
Table \ref{t:twins}.\\ 
$^{b}$ From Mel{\'e}ndez et al.~in prep., fitting solar twin colours as 
a function of $\teff$ and $\feh$.
\end{list}
\end{table*}

In recent years, there has been considerable work in order to
determine the colours of the Sun
\citep[e.g.][]{sekiguchi00,ramirez05b,holmberg06,pasquini08}. One of
the most extensive analysis is that of \citet{holmberg06}: the
remarkably good agreement we have in the optical colours can be understood
from the dependence of these indices on both $\teff$ and $\feh$. The
approximately $100$\,K cooler effective temperature scale adopted by
\citet{holmberg06} favours bluer colours, which are grossly 
compensated to the red by the underestimation of $\sim 0.1$~dex in the GCS
photometric metallicities with respect to spectroscopic ones selected to be 
consistent with our temperature scale \citep[][]{holmberg09}. 
Our $B-V=0.641$ is also in very good
agreement with the $B-V=0.649\pm0.016$ found studying solar twins in
M67 \citep{pasquini08}. For this cluster, using our colour-temperature
relation to compare $V-K_S$ photometry with theoretical isochrones
shows remarkably good agreement (Vandenberg, private communication). 

Infrared indices derived inverting Eq.~(\ref{eq:ct}) depend almost
exclusively on the adopted $\teff$ scale, which is responsible for our much 
redder colours than those of \citet{holmberg06}. Our $V-$ $J$, $H$ and $K_S$ 
are in good agreement with those reported in \citet{rieke08} and obtained from
solar-type stars or computed convolving various solar spectra with the
appropriate filter curves and using their revised absolute physical
calibration.  

The empirical colours in Table \ref{tab:sun} are also in agreement
with the synthetic ones, computed using the same zero points and
absolute calibration for Vega used in the IRFM to derive our $\teff$
scale. Therefore, the uncertainty in the zero points used to generate
synthetic colours is at the smallest level possible, yet of the order
of $0.01$~mag.~(Section \ref{tuning}), allowing us to address the
reliability of the models at this level of precision. 
While using a theoretical spectra of Vega may (partly) compensate model 
inaccuracies in the process of setting the zero points, the approach adopted 
here allows us to focus on the quality of the solar synthetic spectra. 
The agreement is remarkable, on the order of $0.01$ mag.~and never exceeding 
$0.02$, which is also of the same size of the difference between those
synthetic models.   

\section{Conclusions}\label{conclusions}

The primary goal of this work has been to provide a new absolute effective 
temperature scale. An unprecedented accuracy of few tens of Kelvin in the 
zero point of our scale has been achieved using a sample of solar twins. 
For these stars the high degree of
resemblance to the Sun has been determined entirely model
independently, without any prior assumption on their physical
parameters, most importantly $\teff$.
Notice that by calibrating our results via solar twins we are entirely unaffected 
from possible issues and uncertainties related to Vega. Nonetheless, we 
regard as comforting that our findings are in close agreement with the latest 
absolute fluxes \citep{cohen03,bohlin07,rieke08}. We further took advantage of 
such a promising situation by fine-tuning the adopted fluxes so as to improve 
the consistency of the effective temperatures determined from each band used 
in the IRFM. This methodology gives us confidence that the
stellar parameters we determined are well calibrated not only around the
solar value, but over a wide range in $\teff$ and $\feh$. 
Notice that the IRFM is little model dependent and certainly not at the solar 
value because of our calibration procedure. Small spurious trends arising 
from the 
adopted library at different temperatures and metallicities can not be 
entirely ruled out, but should be small.
Though the zero point of our new $\teff$ scale is entirely set by solar twins, 
it agrees within few degrees with independent verifications conducted via 
interferometric angular diameters and HST spectrophotometry in the metal-poor 
and -rich regimes.

In the process of establishing the zero point of the effective temperature 
scale via IRFM, we nailed down the differences with respect to other 
implementations of the same technique. We have used two independent IRFM 
versions
to study the discrepancies among various temperature scales that appeared in
literature over the years and proved that the absolute calibration of the 
photometric systems used was responsible for explaining most of the 
differences. At solar temperatures and metallicities 
the long-standing dichotomy between photometric and spectroscopic 
$\teff$ is easily explained once it is understood that the IRFM can 
in principle accommodate any temperature scale since its zero point 
depends on the absolute calibration of the photometry adopted. 
The main goal of the present paper has been exactly to tackle this issue 
using the best constraint available to date. 

The improved 
bolometric fluxes determined for metal-poor stars have also been used
to put on firmer ground the temperature scale in this rather unexplored
regime. For metallicities typical of halo stars our $\teff$ scale is
roughly $100$\,K hotter than those determined from the Balmer lines
and $200$\,K hotter than those obtained from the excitation
equilibrium. While spectroscopic effective temperature determinations
have considerable model dependence and are degenerate with
other stellar parameters (namely $\logg$ and $\feh$), the IRFM offers a
powerful alternative, free from any of the above limitations. 
However, relying on the photometry, the IRFM is influenced by reddening, 
which becomes a considerable source of uncertainty when
targeting objects outside of the local bubble. For our sample of
metal-poor stars we have been cautious in determining reddening
as best we could. Our improved determination of $E(B-V)$ also explain the 
remaining discrepancies with other $\teff$ scales. 
We think the effective temperatures determined for our sample of stars
will serve to better calibrate spectroscopic $\teff$ 
determinations. This will be
particularly relevant when large spectroscopic surveys targeting
different stellar populations in the Galaxy start operating:
support from the existing or forthcoming photometric surveys will be
possible only if reddening will be determined on a star-by-star
basis. We feel this will not be possible in many cases and stellar parameters
will have to rely on spectroscopy only.

Based on our sample of dwarfs and subgiants, a set of homogeneously
calibrated colours versus temperatures, bolometric fluxes and angular
diameters have also been determined. A number of problems of interest
to stellar and Galactic Chemical evolution depend on the assumption
made in these relations and our results will permit those problems to
be tackled with greater confidence.

\begin{acknowledgements}
We thank Ana Garc\'ia-P\'erez for preliminary results on HE0233-0343 and interesting discussion on excitation temperatures in metal-poor stars. We are also indebted to Maria Bergemann and Andreas Korn for many insights on determining effective temperatures from Balmer lines. Martin Cohen is acknowledged for useful correspondence on the absolute calibration and Gerard van Belle for enlightening discussions on interferometry at various times. Don Vandenberg is kindly acknowledged for useful correspondence and for a preliminary comparison of our temperature scale with M67. We thank an anonymous referee for relevant comments and suggestions that helped to strength the presentation and clarify the results. JM is supported by a Ci\^encia 2007 contract (FCT/MCTES/Portugal and POPH/FSE/EC) and acknowledges financial support from PTDC/CTE-AST/65971/2006 (FCT/Portugal). This research has made use of the General Catalogue of Photometric data operated at the University of Lausanne and the SIMBAD database, operated at CDS, Strasbourg, France. This publication makes use of data products from the Two Micron All Sky Survey, which is a joint project of the University of Massachusetts and the Infrared Processing and Analysis Center/California Institute of Technology, funded by the National Aeronautics and Space Administration and the National Science Foundation.
\end{acknowledgements}

\bibliographystyle{aa}
\bibliography{../../bibs/ir_refs}

\clearpage

\begin{appendix}

\section{Comparing the TCS and 2MASS absolute calibration}

The absolute calibration of \citet{alonso94:absolute_calibration} was obtained 
applying the IRFM to a sample of stars for which direct measurements of 
angular diameters were available. Because of the difficulties involved in 
achieving milli-arcsecond resolution, that sample was almost entirely 
composed of giants with angular diameters measured via Lunar Occultations and 
Michelson Interferometry ($\teff < 5000$~K) or Intensity Interferometry 
($\teff > 6000$~K). One of the intriguing results of that analysis was the 
impossibility of setting the same zero point of the absolute calibration 
using angular diameters measured by Lunar Occultations 
\citep{white87,ridgway80} and Michelson Interferometry 
\citep{hutter89,dbr87,mozurkewich91} with those measured by Intensity 
Interferometry \citep{hanbury74}. The absolute calibration (in the Johnson 
system) proposed by \citet{alonso94:absolute_calibration} is a weighted 
average from their table 10 and it is interesting to notice that the one 
derived from Intensity Interferometry alone is $4.8$ ($J$) $1.3$ ($H$) and 
$4.0$ ($K$) percent lower than the averaged, proposed one. As we have 
discuss throughout the paper\footnote{We have verified using our IRFM 
implementation that a 1\% increase in infrared fluxes correspond to a decrease 
of 20~K in $\teff$.}, lower infrared fluxes support higher $\teff$ 
(in this case, the average difference in Johnson system would be $3.4$\% 
supporting $\teff$ approximately hotter by $70$~K), 
so it is not surprising our effective temperature scale provides good 
agreement with interferometric measurements despite being considerably hotter 
than most of the previous IRFM analyses. 

\begin{figure*}
\includegraphics[width=1.0\textwidth]{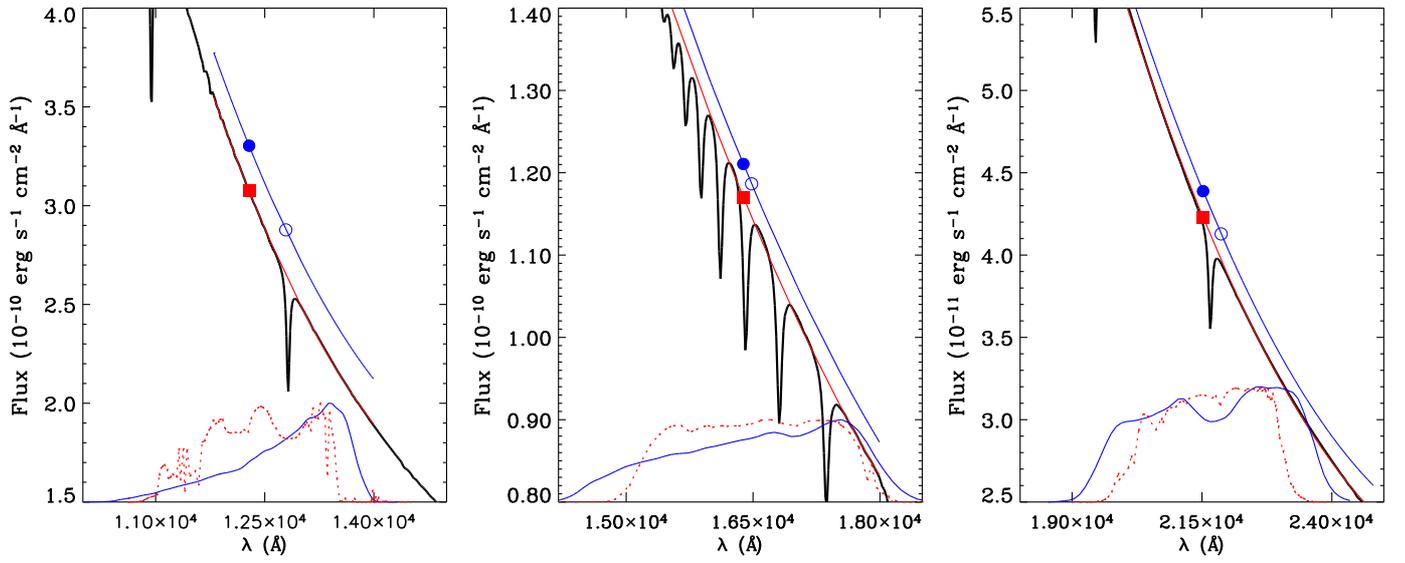}
\caption{Comparison between $\mathcal{F}_{\rm{eff}}$ in 2MASS (filled squares) 
and the TCS before (open circles) and after (filled circles) correcting for 
the same $\lambda_{\rm{eff}}$. The correction has been done shifting the TCS 
values along the continuum of Vega, obtained by fitting a second order 
polynomial to the observed spectral energy distribution 
\citep[from][]{bohlin07}. Overplotted for comparison are the 2MASS (dotted 
lines) and TCS (continuous lines) filter transmission curves.}
\label{f:absflx}
\end{figure*}

To gauge a further insight into the problem, here we directly compare the TCS 
\citep[given in][]{alonso94:photometry} and 
2MASS absolute calibration. Such an exercise, however is not 
straightforward since the absolute calibration in different photometric system 
is obtained using different filter transmission curves and therefore is associated to different effective wavelengths. In addition, for the sake 
of the IRFM, in any given band $\zeta$, it is the composite effect of Vega's 
magnitudes and fluxes which matters, i.e.\ 
$\mathcal{F}_{\zeta} 10^{0.4 \rm{m}_{\zeta}}$. 
Therefore, for a meaningful comparison we need to refer everything to a 
common wavelength, the 2MASS one being the natural choice in this case. 
This is done in Table \ref{2MASSTCS} by computing $\mathcal{F}_{\rm{eff}}$ 
i.e\ the composite effect of magnitudes and fluxes shifted to the 2MASS 
$\lambda_{\rm{eff}}$ in the case of TCS (Figure \ref{f:absflx}).

\begin{table}
\centering
\caption{Characteristic parameters of the 2MASS and TCS photometric systems. 
Wavelengths are in \AA\ and the Vega's monochromatic absolute fluxes in 
$\rm{erg\,cm^{-2}\,s^{-1}}\,\AA^{-1}$.}
\label{2MASSTCS}
\begin{tabular}{lcccccc}
\hline\hline
	                    &   2MASS  &  TCS     &   $\Delta(\%)$  \\
\hline   
$\lambda_{\rm{eff}}$         &  $12285$  & $12790$  &                \\
$\rm{m}_{J}$                &  $-0.001$ & $-0.013$ &                \\
$\mathcal{F}_{J}$           &  $3.079$  & $2.912$  &                \\
$\mathcal{F}_{\rm{eff}}$     &  $3.076$  & $3.303$  &   $-6.9\%$     \\
\hline
$\lambda_{\rm{eff}}$         &  $16385$  & $16483$  &                \\
$\rm{m}_{H}$                &  $+0.019$ & $-0.005$ &                \\
$\mathcal{F}_{H}$           &  $1.150$  & $1.192$  &                \\
$\mathcal{F}_{\rm{eff}}$     &  $1.170$  & $1.211$  &   $-3.4\%$     \\
\hline
$\lambda_{\rm{eff}}$         &  $21521$   & $21869$  &               \\
$\rm{m}_{K}$                &  $-0.017$  & $-0.029$ &               \\
$\mathcal{F}_{K}$           &  $0.430$   & $0.426$  &               \\
$\mathcal{F}_{\rm{eff}}$     &  $0.423$   & $0.439$  &  $-3.6\%$     \\
\hline
\end{tabular}
\begin{list}{}{}
\item[] 2MASS effective wavelengths ($\lambda_{\rm{eff}}$), magnitudes and 
fluxes are from \citet{casagrande06}, where the latter has been modified by 
$-1.6, +1.5$ and $+0.3$ percent in $J,H$ and $K_S$ band respectively as 
described in Section \ref{tuning}. The TCS values are from 
\citet{alonso94:photometry}. For the TCS system $\mathcal{F}_{\rm{eff}}$ has 
been computed by shifting the value at the 2MASS effective wavelength (Figure 
\ref{f:absflx}). $\Delta(\%)$ is the percent decrease of the 
TCS $\mathcal{F}_{\rm{eff}}$ needed to match the 2MASS values.
\end{list}
\end{table}

The 2MASS absolute calibration is on the average lower than the TCS by 
$4.6\%$ (a value qualitatively in agreement with the difference in the 
Johnson system discussed above), thus returning $\teff$ on average hotter 
by $\sim 90$~K, and explaining the bulk of the differences discussed in 
Section \ref{comparisonA} when comparing the sample stars directly. Similar 
conclusions can be drawn when comparing with the absolute fluxes and 
magnitudes of Vega used in table 1 of \citet{gonzalez09}. In this case the 
photometric system is the same (2MASS) and one can directly compare 
$\mathcal{F}_{\rm{eff}}$: the difference is $-3.3\%$ ($J$), 
$+1.3\%$ ($H$) and $-2.8\%$ ($K_S$) thus giving an average of $-1.6\%$ which 
correspond to $\sim 30$~K, again in line with the differences discussed in 
Section \ref{comparisonGB}.

\end{appendix}

\Online

\longtabL{8}{
\begin{landscape}
\begin{longtable}{ccccccccccccccccccccc}
\caption{\label{tab:stars} Sample stars$^{a}$.}\\
\hline\hline
 name        &$\logg$& [Fe/H]& $m_{Bol}$&     $\teff$      &      $\theta$      &  $B$   &  $V$   &  $R_C$ & $I_C$  &   $J$  &   $H$  &  $K_S$ & $E(B-V)$\\   
\hline
BD+000444    & 4.70&  $-0.01$&   8.802 &  $4340 \pm   74$ &  $0.254 \pm 0.009$ & 10.732 &  9.532 &  8.796 &  8.149 &  7.260 &  6.692 &  6.500 &  0.000 \\
BD+002058    & 4.19&  $-1.16$&  10.020 &  $6245 \pm   58$ &  $0.070 \pm 0.001$ & 10.652 & 10.199 &  9.890 &  9.570 &  9.165 &  8.899 &  8.842 &  0.017 \\
BD+004470    & 4.61&  $-1.53$&  10.351 &  $6358 \pm   61$ &  $0.058 \pm 0.001$ & 11.040 & 10.450 &  ----- &  ----- &  9.495 &  9.266 &  9.207 &  0.000 \\
BD+012831    & 4.50&  $-1.27$&  10.677 &  $5331 \pm   50$ &  $0.071 \pm 0.001$ & 11.549 & 10.882 &  ----- &  ----- &  9.455 &  9.063 &  8.993 &  0.000 \\
BD+013597    & 4.04&  $-1.87$&  10.809 &  $6289 \pm   63$ &  $0.048 \pm 0.001$ & 11.544 & 11.089 &  ----- &  ----- & 10.005 &  9.718 &  9.665 &  0.045 \\
BD+023375    & 4.12&  $-2.24$&   9.699 &  $6102 \pm   88$ &  $0.085 \pm 0.003$ & 10.414 &  9.944 &  9.635 &  9.297 &  8.799 &  8.537 &  8.498 &  0.022 \\
\hline
\end{longtable}
$^a$ Sample stars used and derived fundamental physical parameters via IRFM. Apparent bolometric magnitudes ($m_{Bol}$) have been computed according to Casagrande et al. (2006), where the absolute bolometric magnitude of the Sun $M_{Bol,\odot}=4.74$. For each star $m_{Bol}$ is obtained using its bolometric flux and effective temperature and therefore it is already corrected for reddening, if present. Notice however that the observed magnitudes given here are not: before computing bolometric corrections, the observed magnitudes should be corrected using the corresponding $E(B-V)$ given here. Errors have been computed as described in the text, without accounting for the uncertainty in $E(B-V)$: changing it by $\pm 0.01$ would affect $\teff$ by approximately $\pm 50$\,K.
\end{landscape}
}

\end{document}